# Analytical investigation of gas production from methane hydrates and the associated heat and mass transfer upon thermal stimulation employing a coaxial wellbore

M. Roostaie[1] and Y. Leonenko[1,2,*]

[1]*Department of Earth & Environmental Sciences*

[2]*Department of Geography and Environmental Management*

*University of Waterloo*
*200 University Ave. W*
*Waterloo, Ontario, Canada N2L 3G1*

*corresponding author, leonenko@uwaterloo.ca, 519-888-4567 (Ex. 32160)

## Abstract

In this study, an analytical approach has been developed to couple the wellbore heating process by hot water circulation and the associated methane hydrate dissociation in the reservoir. A coaxial wellbore is assumed as the heat source where both conduction and convection heat transfers are considered. It consists of an inner tube and an outer structure of casing, gravel, and cement layers. In the reservoir, a similarity solution employing a moving boundary separating the dissociated and undissociated zones is employed to build the analytical solution. Two different operating schemes for water supply into wellbore heat source have been studied: i) from the inner tube; and ii) from the annulus section of the wellbore. Temperature distribution along the wellbore, temperature and pressure distributions in the reservoir, hydrate dissociation rate, and energy efficiency considering various initial and boundary conditions and reservoir properties are evaluated. The two different operating schemes have almost the same results with slightly higher gas production in the case of hot water entry into annulus, which is in direct contact with the reservoir. Increasing the inlet water temperature or decreasing the wellbore pressure increases gas production. Applying them simultaneously results in a greater gas production and energy efficiency. Some of the reservoir's properties, such as porosity, thermal diffusivity, thermal conductivity, and reservoir thickness, have direct relation with the dissociation rate, but the reservoir's permeability and gas viscosity have almost no impact on the process. The wellbore parameters, such as flow rate of hot water, inlet temperature, and wellbore radius except the inner tube radius, have direct impact on the wellbore mean temperature and the associated results in the dissociation process.

## Introduction

Natural gas hydrates are cage-like substances composed of gas molecules trapped inside water molecules by high pressure and low temperature conditions underground in the presence of sufficient water and gas [1-5]. These substances represent a significant source of energy due to their unique structure and arrangement. More than 160 volumes of gas can be produced from 1 volume of natural gas hydrate, which is about 150 to 180 times greater than the amount of methane present in an equal volume free gas reservoir [6, 7]. However, an accurate estimation of worldwide natural gas hydrate reservoirs or their specific locations is difficult. Methane hydrates (MH), which are reported to be the most common gas hydrate type in nature, can be found mostly in Canada under thick permafrost in Polar Regions or in deep sediments of the continental shelf regions. The total estimated amount of methane is more than $10^{14}$ $m^3$, which is substantially more than the conventional gas reservoirs in Canada [6]. Gas production from MH reservoirs may become necessary in the near future due to higher gas demand (i.e., local residential gas demand in the impassable Northern regions of Canada), or due to the environmental or industrial hazards. On the other hand, MH dissociation can release methane to the environment if its temperature and pressure are shifted from equilibrium induced by the negative effects of global warming and climate change on the regions containing hydrates [8]. Thus, recently, MH reservoirs have attracted researchers' interest as a potential contributor to global warming [9, 10]. The presence of MH reservoirs is also taken into account as a concern in the oil and gas industry during the exploration process [11].

MH reservoirs found in permafrost regions comprises only a small portion of the world's total reservoirs, most of which are in off-shore regions (approximately 3%), but they are accessible and affordable sites for research activities [12]. The field tests that have been performed on the MH reservoirs located in the Malik site, Mackenzie Delta, Northwest Territories, Canada showed the maximum stability depth of MH around 1.4 km, under which the specific temperature and pressure equilibrium conditions of MH cannot be met [13, 14]. MH occurrence is not consistent vertically through a specific range of depth, but, the total overall thickness of MH in the Northwest Territories of Canada is estimated to be around 80 m, with the total area of approximately 125000 $km^2$ [6]. But, more investigations are also required before the practical gas production and extraction occur from such reservoirs.

Main methods applied to dissociate MH are as follows [4, 15]: i) depressurizing MH reservoirs under the equilibrium pressure [16-20]; ii) thermally stimulating the reservoirs by increasing the MH temperature beyond its equilibrium [21-23]; iii) coupling depressurization with thermal stimulation [24-26]; iv) injecting inhibitors, such as methanol to induce instability in MH [27-29]; and v) replacing methane by $CO_2$ in MH reservoirs [30-35]. It should be mentioned that depressurization in conjunction with thermal stimulation is reported to be the efficient than one of the above methods alone [36]. For example, Liang et al. [37] designed and implemented experiments on MH dissociation upon depressurization, electrical heating, and huff and puff using a single vertical well and reported that the combination of depressurization with electrical heating resulted in a better gas production. Nair et al. [38] through experimental investigation on gas production from hydrate bearing clayey sediments underlying a free gas zone revealed that depressurization in conjunction with thermal stimulation has higher efficiency compared to that of employing those methods lonely. Wang et al. [25] by employing a pilot-scale hydrate simulator reported that depressurization and its combination with thermal stimulation are the optimum method respectively for gas-saturated and water-saturated reservoirs. Minagawa et al. [39] experimentally proved that depressurization in conjunction with electrical heating induce higher gas production in MH sediments saturated with NaCl electrolyte solution. Wang et al. [24] by employing an experimental cubic hydrate simulator revealed that thermal stimulation in conjunction with depressurization resulted in higher energy efficiency. Jin et al. [40] numerically investigated gas production from MH reservoirs using horizontal wells in Shenhu area, South China Sea. They reported that hot water injection to the reservoir and depressurizing simultaneously improved gas production from MH dissociation.

Many mathematical studies of hydrate dissociation employing the above methods have been performed so far and can be categorized into the following: i) analytical solutions, which provide fast answers with a good mechanistic prospective of the phenomena; and ii) numerical solutions, which are comprehensive but complicated by a number of assumptions. Analytical solutions are universal, which means that they return the exact solution to problems by direct substitution of variables [41]. They appear in symbolic form and facilitate performance of parametric studies to determine the effect of each parameter of individual variable on the outcome. If the effect of a term dominates the equation, the other terms may be dropped to simplify the equation. It makes a rather inexact solution, but completely sufficient as long as the error is quantifiable and in a limited range [41, 42]. On the other hand, numerical solutions are not universal, meaning that they return an approximate solution to the problem and they cannot assess the exact effect of parameters or dependent variables on the outcome [43]. For example, in complex numerical solutions, in which the equation consists of many different independent variables and parameters that also change over time, it is almost impossible to quantify the effect of each parameter on the outcome. Numerical solutions contain fewer assumptions compared to those of analytical works. They are mainly suggested for the problems consisting of complicated or nonlinear equations with complex geometries [42, 44, 45]. Thus, due to the simple geometry employed in the present work and the fact that the exact effect of wellbore structure, geometry, and the associated inside thermal and fluidic processes on dissociation has not studied yet, analytical approach is chosen to find the exact solution to the problem in this stage, and numerical approaches may be employed for future investigations with more complex geometries and fewer assumptions.

For instance, a three dimensional (3D) numerical study on MH dissociation upon depressurization was performed in 1982 assuming an MH layer with a free gas zone and involving the effect of conduction heat transfer and gas fluid flow [46]. A few years later, in 1986, this work was extended by taking into account the effect of water flow that is produced during the dissociation [47]. In 1991, a numerical study modelled MH dissociation by depressurization, considering three phases, water, gas, and hydrate, in porous media with gas-water flow and without the effect of heat transfer [48]. Then, other research built on this study considering water-gas flow and convective-conductive heat transfer [49, 50]. Thermal stimulation method was first modelled through numerical works by assuming different media permeabilities and an impermeable moving dissociation boundary that separates the dissociated and undissociated MH zones as dissociation progresses [51]. The dissociation continues until the pressure and temperature falls to the equilibrium. A numerical work, in which finite difference method was employed, considering the effect of heat transfer during depressuriza tion reported how the wellbore pressure affects MH dissociation [52].

Different numerical simulators have been developed to investigate MH dissociation, for example TOUGH2 [53]. New modules have since been introduced, enabling the study of different dissociation methods with different components and up to nine phases in both kinetic and equilibrium models [54]. A numerical work employing TOUGH2 to investigate depressurization and thermal stimulation in determining the gas production potential from MH reservoirs located in the Mackenzie Delta, Northwest Territories, Canada, found higher efficiencies when both methods are used together [54, 55]. The kinetic reaction models avoid under-prediction of recoverable MH, but require more computational effort compared to equilibrium reaction models [56]. Researches using the TOUGH-Fx/HYDRATE simulator reported low gas production from dispersed oceanic MH reservoirs with low hydrate saturation upon depressurization, accompanied by high water production, suggesting that gas production from such reservoirs is not economically feasible [57]. In one of the recent studies on MH dissociation, Liu et al. [58] experimentally investigated the heat transfer process inside the reservoir during dissociation. They reported: i) heat transfer from environment to the reservoir by applying depressurization, however by applying both depressurization and thermal stimulation, heat was distributed from the well to the reservoir; ii) combination of depressurization and wellbore heating improved energy recovery and heat absorption, but it caused heat loss by hydrate sediment and produced water; iii) employing depressurization in the early stages of dissociation before applying wellbore heating improved the dissociation process by producing lower amount of water and consuming lower amount of energy; iv) lower wellbore pressure during depressurization in the combination method is economical because it induced more heat transfer from the environment to the reservoir and lower heat loss from the wellbore heating to the sediments and produced water; and v) lower wellbore pressure in production stage induced lower temperature of the hydrate and increased the resulted heat transfer. Increasingly, MH reservoirs are attracting researchers, who study them deeply through mathematical tools as well as real field or experimental work [59-69].

In 1990, Selim and Sloan [70] employed an analytical one-dimensional (1D) model to investigate MH dissociation upon thermal stimulation. They employed a moving dissociation boundary to separate the dissociated zone from the undissociated one. Although, the produced water was assumed to remain motionless in the pores, the effect of gas convection heat transfer and its flow was taken into account. In 1982, hydrate dissociation upon both thermal stimulation using hot water circulation into reservoir and depressurization was investigated by employing two models (i.e., the frontal-sweep model and the fracture-flow model). The authors reported higher efficiencies with depressurization than with thermal stimulation [71]. Makogon [72] analytically investigated the temperature and pressure distributions in MH reservoirs upon depressurization, considering the throttling effect in the energy equation, and assuming a moving dissociation boundary as in the previous work. This work was continued by including similarity solutions in temperature and pressure calculations as well as the water and gas movement effects [73]. In 2001, Makogon's model [17] was extended by including heat conduction. An analytical study on depressurization method showed that the effect of gas-water (two-phase) flow on MH dissociation is smaller than the effect of heat transfer and the intrinsic kinetics of MH decomposition [74]. Recently, Roostaie and Leonenko [75, 76] employed flat and radial analytical models to investigate MH dissociation upon wellbore heating. They took into account heat transfer from wellbore to the hydrate reservoir through the wellbore thickness.

Apart from the mathematical investigations, several experiments have also been designed and performed on MH dissociation upon different methods. One of the major challenges in those experiments, which significantly affects the outcome of tests, is the size of the setup [77]. For example, the main involving mechanism in the hydrate dissociation in porous media depends on the scale of experiments [77]. This mechanism could be one of the followings: i) heat transfer in the decomposing zone; ii) intrinsic kinetics of hydrate decomposition; or iii) multiphase flow (i.e., gas-water flow) during gas production [74]. Tang et al. [78] reported that the determining factor in core-scale and larger scale experiments (or field works) is respectively the intrinsic kinetics of hydrate decomposition and heat transfer in the decomposing zone. An experimental work by Li et al. [79] employing a 3D cubic hydrate simulator (CHS) showed that MH dissociation upon thermal stimulation progresses by a moving boundary. Li et al. [80] designed experiments by employing two hydrate simulators with different scales to investigate the MH dissociation upon depressurization. They reported that the heat transfer from ambient is the main factor for MH dissociation. They also showed that the size of the reservoir affects the gas production rate and time. Another experimental work revealed that the conduction heat transfer is the main heat transfer mechanism to the dissociating zone [81]. Li et al. [82] employed a Cuboid Pressure Vessel to investigate MH dissociation upon electrical heating and depressurization. They reported that the hydrate dissociated completely in the experimental setup in a shorter time upon electrical heating compared to that upon depressurization. A 3D Pilot-Scale Hydrate Simulator (PHS) was designed by Wang et al. [77] to investigate MH dissociation upon depressurization. They simulated MH dissociation below the quadruple point in the sandy sediment using PHS and reported ice formation in pores of the media, which increased the dissociation rate. Another experimental work employing a pilot-scale hydrate simulator reported that depressurization in conjunction with thermal stimulation is the most efficient method for the dissociation of water-saturated MH [25]. Recently, an experimental work studied depressurization, thermal stimulation, and depressurization in conjunction with thermal stimulation methods in MH dissociation this work, which was supported by an analytical model, also showed that employing depressurization and thermal stimulation simultaneously provided higher efficiency [83]. Li et al. [84] experimentally showed higher gas production from MH dissociation upon depressurization and electrical heating by designing a Cuboid Pressure Vessel. Schicks et al. [85] designed and developed a large laboratory reservoir simulator and investigated the potential use of in-situ combustion in hydrate dissociation. They also used the experimental data to validate their numerical one. Davies et al. [86] experimentally and numerically investigated hydrate dissociation upon depressurization and electrical heating.

During the previous decades, several investigations have been performed on gas production from MH dissociation and the associated changes in the reservoir [34, 87-92]. However, no analytical study has considered the effect of wellbore geometry and associated structure-generally consisting of casing, cement, and gravel [93-95]- also despite various experimental and mathematical studies on wellbore heating by electrical heating, none of the previous works studied the effect of wellbore heating by hot water circulation (i.e., heat pump) on MH dissociation so far [37, 39, 63, 84, 86, 96-98]. Our recent works [75, 76], which were verified against previous numerical and experimental works, showed that the wellbore structure (i.e., casing, gravel, and cement external layers) affects the dissociation upon wellbore heating. In those investigations [75, 76], we employed two flat and radial geometries, in which the temperature of the heat source was assumed to be constant implying that the process involved in heating the heat source and the associated inside structure of the wellbore were not considered. In investigations related to geothermal energy, ground-source heat pumps are employed as a heat source/sink to derive heat transfer process to/from underground [99-103]. Heat pumps come in two types [104]: i) Coaxial boreholes heat exchangers; and ii) U-tube heat exchangers. These heat exchangers do not inject water directly into reservoirs, but rather they indirectly transfer heat through both their own and the wellbore's structure. This study aims to expand our previous works [75, 76] by considering a cylindrical coaxial wellbore as the heat source with inside hot water circulation to heat that up and coupling the inside hot water circulation to heat that up (e.g., heat source operating schemes, performance, and fluidic and thermal processes inside the wellbore) to the dissociation process and the associated interactions inside the reservoir. The present work focuses on the coaxial boreholes heat exchangers because their associated design has been used and investigated since decades ago [105-107]. They also have higher efficiencies in terms of heat transfer to the reservoir compared to those of U-tube heat exchangers [108-110]. It should also be noted that employing radial coordinates and the mechanism of heating the wellbore and its effects on dissociation in the previous analytical studies of MH dissociation upon thermal stimulation have not been treated in detail. These assumptions make the outcomes closer to the real operational conditions and facilitate further studies and optimizations on this process.

The purpose of the present work is to develop radial analytical models of MH dissociation upon thermal stimulation, assuming an infinite hydrate reservoir in the radial coordinates and considering a wellbore heat source with three main completion layers- the casing, cement, and gravel-as its external structure. A coaxial heat exchanger with an inside tube and a surrounding annulus is assumed to model hot water injection in the wellbore and heat transfer to the reservoir. The energy efficiency, gas production, and temperature and pressure distributions assuming different heat exchanger and reservoir characteristics are calculated and compared with the previous studies. The results of this study, for the first time, provides an important opportunity to advance the understanding of the effect of wellbore structure, composition, and its associated heating mechanism on the gas production from MH reservoirs upon thermal stimulation.

## Materials and methods

A schematic of the dissociation process inside the reservoir is presented in Figure 1a, which is similar to that of our previous work [76] except the wellbore structure (grey region). Zone II is composed of undissociated hydrate at a temperature of $T_0$ at infinity ($r \to \infty$), which increases to converge to the dissociation temperature ($T_s(t)$) as it approaches the moving dissociation boundary ($r=S(t)$), denoted by the dashed circle. Zone I represents the dissociated region, with its different temperatures and pressures at various locations ($r$) and times ($t$). The grey ring in the middle of Figure 1a shows the supporting structure of the wellbore. A top view of the wellbore is shown in Figure 1b. The wellbore considered in this study has coaxial configuration with pipe-in-pipe geometry [105, 106]. Hot water is supplied to the wellbore either through the inner tube or the annulus between the inner and outer tubes, it is then extracted from the other path (Figure 1b). Heat transfer to the reservoir occurs through the borehole's external wall. The borehole's external layers, shown in Figure 1c, appear in cross section area of the borehole wall in Figure 1b. These layers (from inner to outer) are the 1) inner tube wall; 2) inside casing; 3) gravel; 4) outside casing; and 5) cement, and the associated radii are shown in Figure 1c. The following basic steps represent the MH dissociation process in this modelling: i) before dissociation begins, MH is assumed to be in equilibrium condition at a temperature $T_0$ in the reservoir pores; ii) at time $t = 0$, hot water is injected into wellbore with a constant flow rate and temperature, inducing heat transfer to the reservoir; iii) when the temperature at the external wall of the borehole reaches the dissociation temperature of MH, dissociation begins, followed by a sharp moving boundary surface separating the dissociated zone (the water and gas produced in Zone I) from the undissociated zone (the undissociated MH in Zone II).

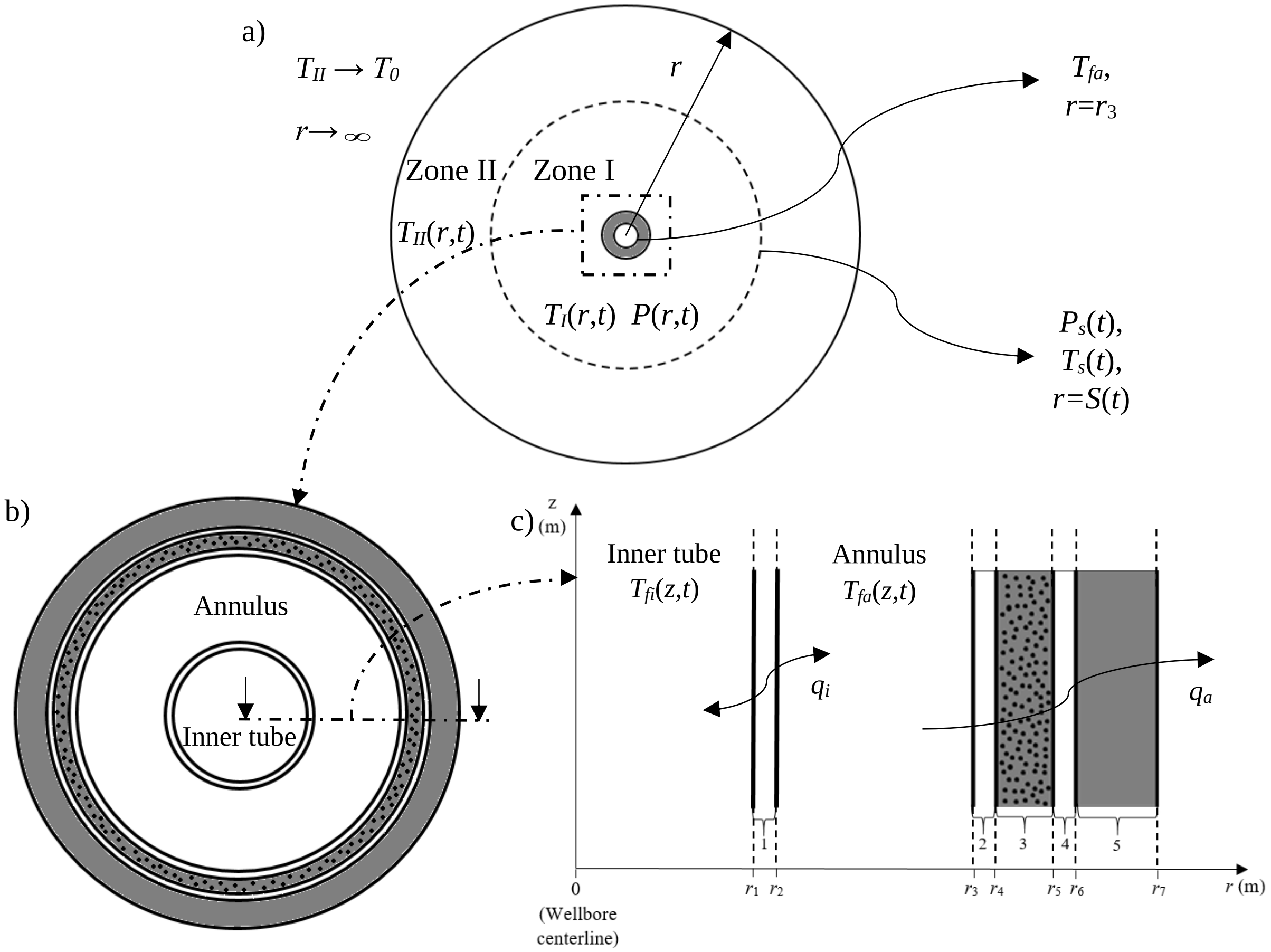

Figure 1. a) Schematic of the infinite radial model of hydrate dissociation considered in this study. The dissociation interface is identified by the dashed circle, and the grey region shows the well thickness. b) Magnified image of the borehole top view. c) Borehole structure with borehole center line denoted by a dashed line, and the various layers: 1) Inner tube wall; 2) Inside casing; 3) Gravel; 4) Outside casing; and 5) Cement.

As mentioned previously, two operating conditions are considered for the heat source: i) hot water injection into the inner tube and extraction from the annulus; and ii) hot water injection into the annulus and extraction from the inner tube. The flow rate and the inlet temperature are constant in models. However, the temperature distribution of the fluid along the inner tube ($T_{fi}$) and the annulus ($T_{fa}$) is not constant, and changes over time due to the heat transfer to the outside via conduction and the convection heat transfer of the fluid inside the wellbore.

The heat transfer rate inside the wellbore (e.g., between the annulus and inner tube) and to the reservoir are represented respectively by $q_i$ and $q_a$ (Figure 1c). The former ($q_i$) consists of three heat transfer components: i) convection heat transfer in the inner tube between the water flow and the inner surface of the tube; ii) conduction heat transfer through the inner tube thickness to the annulus; and iii) convection heat transfer between the outer surface of the inner tube and the water flow in the annulus. The latter ($q_a$) consists of: i) convection heat transfer between the water flow and the outer surface of the annulus; and ii) conduction heat transfer through the external wall of the wellbore (cement, casing, and gravel layers). For each of the heat transfer components of $q_i$ and $q_a$, the thermal resistance is defined in the following based on the heat transfer calculation of the wellbores presented by Hellstrom [111] in 1992.

Convective heat transfer resistance (m.K/W) between the inner tube surface ($r$=$r_1$) and the water flow inside the tube is defined as follows:

$$R_{cii} = \frac{1}{\pi k_f Nu_{ii}} \tag{1}$$

where $R_{cii}$ is the convective heat transfer resistance (m.K/W), $k_f$ is the fluid thermal conductivity (W/m.K), and $Nu_{ii}$ is the Nusselt's number of the fluid flow in the inner tube. Different formulas have been suggested for Nusselt's number in previous literature with various degrees of accuracy [112-119]. Among them, Gnielinski [119] reported a suitable and accurate correlation for coaxial heat exchangers considering fluid flow in the annulus [113, 120-122]. Nusselt's number of fluid flow in the inner tube is defined as follows based on the Gnielinski formula [119]:

$$Nu_{ii} = \frac{(f\ /2)(\mathrm{Re}_i - 1000)\,\mathrm{Pr}}{1 + 12.7(f\ /2)^{1/2}(\mathrm{Pr}^{2/3} - 1)} \tag{2}$$

where $\mathrm{Re}_i$ is the Reynolds number in the inner tube, Pr is the Prandtl number, and $f$ is the friction factor for smooth pipes. $\mathrm{Re}_i$, Pr, and $f$ are calculated respectively as follows:

$$\mathrm{Re}_i = \frac{v_f D}{\nu_f} = \frac{2 r_1 v_f \rho_f}{\mu_f} = \frac{2 \rho_f V_f}{\pi \mu_f r_1} \tag{3}$$

$$\mathrm{Pr} = \frac{\mu_f C_f}{\rho_f k_f} \tag{4}$$

$$f = (1.58 \ln(\mathrm{Re}_i) - 3.28)^{-2} \tag{5}$$

where $D$ is the pipe diameter (m), $v_f$ is the water flow velocity (m/s), $\nu_f$ is the kinematic viscosity ($m^2$/s), $\rho_f$ is the water density (kg/$m^3$), $V_f$ is the water flow rate ($m^3$/s), $r_1$ is the radial distance (m) as shown in Figure 1c, $C_f$ is the fluid volumetric heat capacity (J/($m^3$.K)), and $\mu_f$ is the water dynamic viscosity (Pa.s).

The conduction thermal resistance (m.K/W) of the tube is calculated as follows:

$$R_p = \frac{\ln(r_2 / r_1)}{2\pi k_p} \tag{6}$$

where $r_2$ and $r_1$ are inner tube radii as shown in Figure 1c, and $k_p$ is the inner tube's thermal conductivity (W/(m.k)).

The convection thermal resistance (mK/W) of the heat transfer from the outer surface of the inner tube ($r$=$r_2$) to the fluid flow in the annulus can be represented as follows [111]:

$$R_{coi} = \frac{1}{\pi k_f Nu_{oi}}\left(\frac{1}{r^*} - 1\right) \tag{7}$$

where $Nu_{oi}$ is the Nusselt's number of the fluid flow close to the outer surface of the inner tube, and $r^*$ is the ratio between the inner and outer radii of the annulus $(r_2 / r_3)$. Petuhkov and Roizen [123] derived an expression for the Nusselt's number in the annulus close to the inner tube based on experimental data,

$$Nu_{oi} = \zeta Nu_{pipe} 0.86(r^*)^{-0.16} \tag{8}$$

where $\zeta$ is a constant, equal to 1 in this case, and $Nu_{pipe}$ is the Nusselt's number for turbulent flow in a circular pipe, calculated based on equation 2, using the Prandtl number and friction factor formulas from equations 4 and 5, and a Reynolds number as follows:

$$\mathrm{Re}_a = \frac{v_f d_h}{\nu_f} = \frac{2(r_3 - r_2) v_f \rho_f}{\mu_f} = \frac{2(r_3 - r_2)\rho_f V_f}{\pi(r_3^2 - r_2^2)\mu_f} \tag{9}$$

where $r_3$ and $r_2$ are the radii (m) of the annulus, shown in Figure 1c, and $d_h$ is the hydraulic diameter (m) of the annulus, which can be calculated as follows:

$$d_h = 2(r_3 - r_2) \tag{10}$$

The convection thermal resistance (m.K/W) of water flow and the outer surface of the annulus ($r$=$r_3$) can be calculated based on the following formula [111]:

$$R_{cia} = \frac{1}{\pi k_f Nu_{oa}}\left(1 - r^*\right) \tag{11}$$

where $Nu_{oa}$ is the Nusselt's number of the water flow close to the outer surface of the annulus, which is defined as follows based on the work performed by Petuhkov and Roizen [123]:

$$Nu_{oa} = Nu_{pipe}(1 - 0.14(r^*)^{0.6}) \tag{12}$$

And finally, in order to calculate the conduction thermal resistance of the external wall of the wellbore, the wall can be considered as a homogenous layer by summing the thermal resistances of the cement, casing, and gravel layers together as stated in equation 13:

$$R_w = \frac{\ln(r_4 / r_3)}{2\pi k_p} + \frac{\ln(r_5 / r_4)}{2\pi k_g} + \frac{\ln(r_6 / r_5)}{2\pi k_s} + \frac{\ln(r_7 / r_6)}{2\pi k_c} \tag{13}$$

where $k_x$ stands for thermal conductivity (W/(m.K)) with the *s*, *c*, *p*, and *g* subscripts respectively referring to the steel (casing), cement, inner pipe, and gravel. Figure 1 also shows $r_3$-$r_7$.

Therefore, the total thermal resistances ((m.K)/W) associated with $q_i$ and $q_a$ are as follows:

$$R_1 = R_{cii} + R_p + R_{coi} \tag{14}$$

$$R_2 = R_{cia} + R_w \tag{15}$$

At steady-state condition, the convection heat transfer in the water flow should be equal to the transverse heat transfer between the inner tube and annulus ($q_i$) and between the annulus and wellbore ($q_a$). This expression is presented in equations 16 and 17:

$$\pm C_f V_f \frac{\partial T_{fi}(z,t)}{\partial z} = \underbrace{\frac{T_{fi}(z,t) - T_{fa}(z,t)}{R_1}}_{q_i(z,t)} \tag{16}$$

$$\pm C_f V_f \frac{\partial T_{fa}(z,t)}{\partial z} = \underbrace{\frac{T_{fa}(z,t) - T_I(r_7,t)}{R_2} + \frac{T_{fa}(z,t) - T_{fi}(z,t)}{R_1}}_{q_a(z,t)} \tag{17}$$

where $T_{fi}$ is the water flow temperature (K) in the inner tube, and $T_{fa}$ is the water flow temperature in the annulus (K), and $z$ represents the axial distance along the wellbore, which is in the range of 0 (at the base of the wellbore) to h (top boarder of the hydrate zone). Some notations in equations 16 and 17 should be taken into account: i) the $\pm$ sign refers to the direction of the water flow, with + standing for the water flow in the direction of the z-axis; and ii) in $q_a$ calculation, an average temperature along the outer surface of the external wall of the wellbore ($T_I(r_7)$) is considered. Considering the latter assumption, it should be noted that temperature profile along the outside surface of the wellbore with respect to z-coordinates is calculated ($T_{fi}(z,t)$ and $T_{fa}(z,t)$). Due to the matrix structure in the dissociated zone, which consists of porous media as well as methane gas and water created from the dissociation, both conduction and convection heat transfers exist in this zone. Inconstant temperature along the wellbore surface, especially through the lower-end section, would affect the water flow (produced from dissociation), which tends to accumulate in the bottom of the reservoir due to the earth gravity, through convection heat transfer. However, in the present study, the generated water from the dissociation is assumed to remain motionless in the pores. Thus, the authors decided to investigate the effect of inconstant temperature along the wellbore surface on the dissociation in the future works by considering the water flow generated from the dissociation to avoid unreliable results (caused by considering the inconstant temperature along the wellbore surface and neglecting the water flow inside the reservoir), which make the model away from the real conditions. Furthermore, a full discussion of dependency of the interactions and thermal processes inside MH reservoir on the wellbore's inconstant surface temperature lies beyond the scope of this study. The main contribution of this work is to investigate the impact of wellbore heating process by hot water circulation on MH dissociation and vice versa.

The temperature distribution resulted from the two operation models ($T_{fi}$ and $T_{fa}$) can be obtained by solving equations 16 and 17 and considering two boundary conditions: i) the inlet temperature ($T_i$) is constant; and ii) no heat flux occurs at the base of the wellbore. More details are provided in the supplementary file. Then, the heat transfer rate to the reservoir can be obtained, after which, the subsequent hydrate dissociation will be calculated as described next.

Basically, the heat transfer from the heat source to the reservoir is consumed in two ways: i) by increasing the temperature of the sediments and the water-gas produced in the dissociated zone; and ii) hydrate dissociation at the moving boundary and increasing the temperature of the matrix materials in Zone II close to the moving boundary. The input heat should transfer through the dissociated region, during which, a large part of the input heat will be consumed in the first way mentioned above (i), reducing the rate of hydrate dissociation and the speed of the moving interface as the process continues.

Figure 2 shows the trends of temperature and pressure distribution in the reservoir during dissociation. The temperature and pressure ranges for a specific time in Zone I are respectively $T_s < T_I < T_I(r_7)$ and $P_i < P < P_s$, and in Zone II, the temperature range will be $T_0 < T_{II} < T_s$ with a constant pressure equal the equilibrium pressure of MH. The temperature at the outer surface of the wellbore ($r = r_7$) changes with time due to the heat transfer from the wellbore and temperature changes of the inside fluid, but it is always lower than the temperature on the other side of the wellbore ($T_{fa}$). The temperature and pressure at the moving dissociation boundary are not also constant.

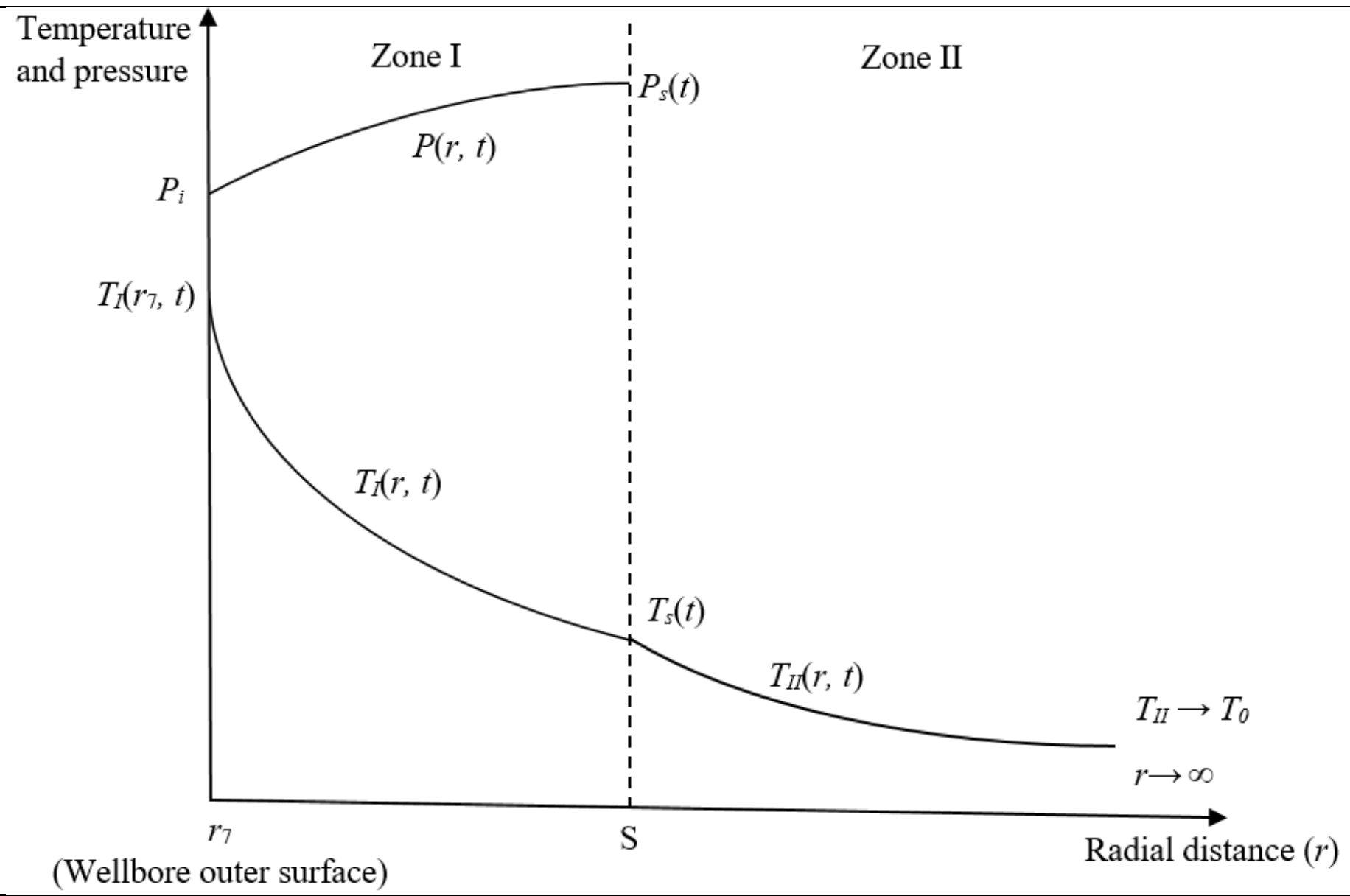

Figure 2. Schematic of pressure and temperature distribution in the reservoir during MH dissociation.

The produced gas will stream towards the heat source according to Darcy's Law, inducing a sudden change in density at the dissociation front due to the gas production. Assumptions made in the dissociation models of the current work are provided in the following in accordance with previous analytical works [23, 70, 75, 76]:

i) MH completely fills the porous media; Well log data determinations and core studies in Malik site, located in Mackenzie Delta, revealed that the hydrate saturation exceeds 80% in at least 10 discrete MH layers [124]. Hence, assuming that the hydrate filled the pores initially is not so far from the real conditions. Furthermore, some of previous analytical studies on MH dissociation assumed that the hydrate filled the porous media completely [23, 70, 75, 76, 125]. Tsimpanogiannis and Lichtner [23] in their analytical study on MH dissociation upon thermal stimulation assumed that the hydrates saturated the pore space completely, which applied an upper limit to the problem and stated that the assumption of gas flow and motionless water restricted the hydrate saturation to the maximum possible values.

ii) thermodynamic equilibrium is applied at the dissociation interface for temperature and pressure;

iii) the water produced during the dissociation remains motionless in the porous media; The volume of generated water from MH dissociation is negligible compared to that of generated gas [6, 7, 125] implying that the convection heat transfer by water has insignificant impact on MH dissociation. Furthermore, previous investigations on MH dissociation showed weak influence of water flow on gas production. For example, Zhao et al. [126] by developing and validating 2D axisymmetric numerical model, showed that the influence of convection heat transfer of the generated water and gas on hydrate dissociation using thermal stimulation method is weak. In another numerical work [127], they reported that the effect of convective heat transfer of water on dissociation upon depressurization is weaker compared to that of the produced gas. Sparrow et al. [128] numerically studied the convection and conduction heat transfers involved in melting of a solid by a vertical tubed-heater. Their analysis showed that for conduction only, the heat transfer rate decreased consistently. However, convection heat transfer by water movement caused a decrease in the heat transfer rate in early stages until attaining a minimum, then raised to a maximum, and finally decreased consistently. This behavior is due to weak convection heat transfer at early stages, then as the melt region grows, convection heat transfer becomes stronger and increased the overall heat transfer rate, however as the melt region thickness grows sufficiently, a boundary layer mechanism induces no direct conduction heat transfer through the melt region, and only convection heat transfer by circulating fluid caused consistent drop in the heat transfer rate. They also revealed that the thickness of the melt region varies along the vertical tube, with maximum thickness at the top due to the convection heat transfer by water movement. Hence, due to the weak effect of water movement on heat transfer and eventually the dissociation process, assuming motionless water produced from dissociation causes negligible error in the final outcome. However, the convection heat transfer by gas is taken into account. Furthermore, in analytical approaches, if effect of a term dominates, the other terms may be ignored to reduce the complexity of the solution and the computation costs [41, 42]. It should be mentioned that previous analytical studies on MH dissociation did not consider convection heat transfer by dissociated water [23, 70, 75, 76, 125, 129-131];

iv) the reservoir is assumed to be infinite; The idea of heat transfer from environment to the reservoir occurs in a finite MH reservoir model. Wang et al. [83] developed an analytical model on MH dissociation upon thermal stimulation based on their experimental work. Thus, they assumed a finite MH reservoir in their analytical study, in which there is heat transfer from environment to the reservoir. They reported that heat transfer from the ambient to the reservoir mainly affects dissociation upon depressurization. Song et al. [132] also through experimental work, reported that heat transfer from environment to the reservoir mainly affect the dissociation process upon depressurization. Feng et al. [4] through experimental study on MH dissociation upon thermal stimulation showed that the forced convection induced by hot water injection into the reservoir increases the effect of heat transfer from environment to the reservoir. Schicks et al. [133] insulated their experimental setup to reduce the heat transfer between the environment and the reservoir in order to make their experimental conditions closer to those of real field tests. Liu et al. [58] through their experimental work on MH dissociation showed that heat transfer from the ambient to the reservoir improved the process upon depressurization. Liang et al. [37] experimentally revealed that heat transfer from environment to the reservoir improved dissociation upon depressurization, on the other hand, they reported that heat loss to the environment upon wellbore heating method is small. Holder et al. [134] in their mathematical study on MH dissociation upon steam injection into the reservoir, assumed a constant percentage of heat loss to the environment by fluid flow. Analytical work by Selim and Sloan [70] and our recent studies on thermal stimulation [75, 76] did not consider the heat transfer from environment to the reservoir. It should be noted that in thermal stimulation method, there could be heat loss to over burden units induced by water and gas convection heat transfer, but there is no heat transfer from those units to the reservoir because the temperature of the dissociated zone is higher than that of those units, and the temperature far inside the hydrate zone is equal to that of those units.

v) thermophysical properties of the phases remain constant during dissociation; v) the produced gas is modelled as an ideal gas;

vi) the produced gas instantaneously reaches the same temperature as that of local sediments;

vii) no viscous dissipation or inertial effect exists.

The following formulas represent the basic equations and solution procedure used in exploring dissociation, which are similar to those of our previous work [76], while considering heat transfer from the wellbore.

The continuity equation of gas in Zone I is:

$$\phi\left(\frac{\partial \rho_g}{\partial t}\right)+\left(\frac{\partial \rho_g v_g}{\partial r}\right)=0,\ t>0 \tag{18}$$

where $\rho_g$ is gas density (kg/m$^3$), $\phi$ is reservoir porosity, $r$ is the radial distance (m), and $v_g$ is gas velocity (m/s). The gas velocity ($v_g$) in Zone I is calculated as follows by employing Darcy's Law:

$$v_g=-\left(\frac{k}{\mu}\right)\left(\frac{\partial p}{\partial r}\right),\ t>0 \tag{19}$$

where $k$ is gas effective permeability (μm$^2$), $\mu$ is gas dynamic viscosity (Pa.s), and $P$ is gas pressure (Pa).

Equations 20 and 21 show the energy balance in Zones I and II, respectively:

$$\rho_I C_{pI}\frac{\partial T_I}{\partial t}+\frac{\partial \rho_g C_{pg} v_g T_I}{\partial r}=k_1\frac{\partial}{\partial r}r\frac{\partial T_I}{\partial r},\ t>0,\ r_7<r<S \tag{20}$$

$$\frac{\partial T_{II}}{\partial t}=\frac{\alpha_{II}}{r}\frac{\partial}{\partial r}r\frac{\partial T_{II}}{\partial r},\ t>0,\ S<r \tag{21}$$

where $\rho_I$ is the density (kg/m$^3$) of the matrix in Zone I, $\rho_g$ is gas density (kg/m$^3$), $C_{pI}$ is the specific heat capacity (J/(kg.K)) of the matrix in Zone I, $C_{pg}$ is the specific heat capacity of gas (J/(kg.K)), $T_I$ is the temperature (K) of the matrix in Zone I, $T_{II}$ is the temperature (K) of the matrix in Zone II, $\alpha_{II}$ is the thermal diffusivity (m$^2$/s) of the matrix in Zone II, and $k_I$ is the thermal conductivity (W/(m.K)) of the matrix in Zone I.

The gas density in Zone I, based on the ideal gas law, can be calculated by the following equation:

$$\rho_g=\frac{mP}{RT_I},\ r_7<r<S, t>0 \tag{22}$$

where $m$ is the gas's molecular mass (kg/mol), and $R$ is the universal gas constant (J/(mol.K)). The above equations are representative of the fundamental concept of the MH dissociation process employed in the present work. The initial and boundary conditions are: i) temperature and pressure at the inner wall of the wellbore ($r_3$) are known at each time step due to the calculations mentioned earlier for temperature distribution inside the wellbore; and ii) no pressure drop is assumed in the wellbore, and the pressure at the outer surface of the well ($r_7$) is $P_i$.

The heat transfer from the wellbore's external wall to the reservoir is stated by Equation 23:

$$-k_I A_w \frac{\partial T_I}{\partial r} = \frac{(T_{fa} - T_I)}{R_2}, r = r_7, t > 0 \qquad (23)$$

where $A_w$ is the wellbore surface area (m$^2$), $k_I$ is the thermal conductivity of Zone I (W/(m.K)), $r$ is the radius (m). It should be noted that various structures and geometries for wellbores exist [95, 135, 136], and the wellbore structure proposed in the present work is assumed as a general model based on previous works [93, 94, 100, 137].

There is thermodynamic connection between the temperature and pressure at the moving dissociation interface, represented by the Antoine Equation as follows:

$$P_s = \exp(A_a - B_a / T_s), \; r = S, t > 0 \qquad (24)$$

where $P_s$ and $T_s$ are respectively the pressure (Pa) and temperature (K) at the moving dissociation interface, $A_a$ and $B_a$ are constants.

Equations 25, 26, and 27 respectively represent the mass and energy balances at the dissociation interface and the MH dissociation heat [70] with the associated initial and boundary conditions represented through equations 28-30:

$$F_{gH} \sigma_H (\frac{dS}{dt}) + \rho_g v_g = 0, \; r = S, t > 0 \qquad (25)$$

$$k_{II} \frac{\partial T_{II}}{\partial r} - k_I \frac{\partial T_I}{\partial r} = \phi \rho_H Q_{Hd} \frac{dS}{dt}, \; r = S, t > 0 \qquad (26)$$

$$Q_{Hd} = c + dT_s \qquad (27)$$

$$T_I = T_{II} = T_s(t), \; r = S, t > 0 \quad (28)$$

$$T_{II} = T_0, \; r \to \infty, t > 0$$

(29)

$$T_{II} = T_0, r_7 < r < \infty, t = 0, \; S = 0 \quad (30)$$

where $F_{gH}$ is the mass ratio of the methane gas trapped inside the MH to the total mass of hydrate and is assumed to be equal to 0.1265 kg $CH_4$/kg hydrate [70], $\rho_H$ is the hydrate density (kg/m$^3$), $k_{II}$ is the thermal conductivity (W/(m.K)) of Zone II, $Q_{Hd}$ is MH dissociation heat (J/kg), and $c$ and $d$ are constants.

Equations 18, 20, and 24 are simplified by employing equations 19 and 22 to eliminate the gas velocity and density, provided through the following equations 31-33:

$$\phi \frac{\partial}{\partial t}\left(\frac{P}{T_I}\right) - \frac{k}{\mu}\frac{\partial}{\partial r}\left(\frac{P}{T_I}\frac{\partial P}{\partial r}\right) = 0 \quad (31)$$

$$\rho_I C_{pI} \frac{\partial T_I}{\partial t} + \frac{kmC_{pg}}{\mu R}\frac{\partial}{\partial r}\left(P\frac{\partial P}{\partial r}\right) = \frac{k_1}{r}\frac{\partial}{\partial r} r \frac{\partial T_I}{\partial r}, \; t > 0, \text{ in Zone I} \quad (32)$$

$$F_{gH}\phi\sigma_H\left(\frac{dS}{dt}\right) - \frac{kmP}{\mu R T_I}\frac{\partial P}{\partial r} = 0, \; r = S, t > 0 \quad (33)$$

To solve the abovementioned equations, the similarity solution is employed, using a dimensionless parameter relating the movement of the dissociation interface to the square root of time ($t^{1/2}$), as represented by a non-dimensional parameter shown in equation 34. This method, first introduced by Neumann [138, 139], also satisfies the initial and boundary conditions.

$$\lambda = \frac{r}{\sqrt{4\alpha_{II} t}} \quad (34)$$

On the moving dissociation interface, equation 20 becomes:

$$\beta = \frac{S}{\sqrt{4\alpha_{II} t}}, \; r = S, t > 0 \quad (35)$$

And, on the outer surface of the wellbore:

$$\lambda_{os} = \frac{r_7}{\sqrt{4\alpha_{II} t}}, \; r = r_7, t > 0 \tag{36}$$

Therefore, by using equations 34-36, the previous equations are simplified and transformed, as presented in the supplementary information in detail.

An exponential integral (Ei) function is employed to find the solution to the temperature distribution in the reservoir during dissociation as described in equations 37 and 38. This solution has also been recommended in previous works [138, 139].

$$T_I = -AEi(-(a\lambda + b)^2) + AEi(-b^2) + B \tag{37}$$

$$T_{II} = T_0 + CEi(-\lambda^2) \tag{38}$$

The *A*, *B*, *C*, *a*, and *b* constants are defined in the supplementary information. The parameter "b" in equations S17 and S25 defines the convection heat transfer for gas by taking into account the specific heat capacity of gas. The gas produced from dissociation will absorb some of the transferred heat from the wellbore to the dissociation interface. This definition is in line with previous analytical studies [23, 70, 75, 76, 129].

Then, the pressure distribution in Zone I can be calculated from equation S12 as follows:

$$P(\lambda) = \left( P_0^2 + \frac{4F_{gH}\phi\rho_H\alpha_{II}\mu R\beta}{km} \int T_I d\lambda \right)^{1/2} \tag{39}$$

By replacing $T_I$ according to equation 37, the pressure distributions in zone I will be calculated as shown in equation 40:

$$P = \left( P_0^2 + L(\beta)(K(\beta)\lambda - AN(\lambda) - (K(\beta)\lambda_{os} - AN(\lambda_{os}))) \right)^{1/2}, \tag{40}$$

where $L(\beta)$, $N(\lambda)$, and $K(\beta)$ are defined in the supplementary information. It should be noted that the obtained expressions for temperature and pressure distributions can satisfy the basic equations and boundary conditions, mentioned earlier, by direct substitution.

Heat flux (J/(m$^2$. s)) from the wellbore to the reservoir as a function of time can be obtained from the following formula:

$$u_r = -k_I \frac{\partial T_I}{\partial r}, \; r = r_7 \tag{41}$$

Equation 41 can be transformed into an equal equation according to equation 20 (supplementary information).

By integrating Equation 41, the total heat input into the reservoir from the wellbore (J/m$^2$) up to time *t* can be calculated as in the following equation:

$$Q_{rt} = -k_I \int_0^t \frac{\partial T_I}{\partial r} dt \ , \ r = r_7 \tag{42}$$

The total volume of gas produced up to time *t* can be calculated as follows at the standard temperature and pressure (STP) of dry gas:

$$V_{rp} = \frac{n_{rt} R T_{STP}}{P_{STP}} \tag{43}$$

where $V_{rp}$ and $n_{rt}$ are respectively the total volume (m$^3$/m$^2$) and total moles (mole/m$^2$) of produced gas per surface area of the moving interface up to time *t* with $T_{STP}$ and $P_{STP}$ respectively as the temperature and pressure of dry gas at STP conditions. Further details of the calculation of the total volume of produced gas are provided in the supplementary information.

In order to assess the efficiency of gas production during a dissociation process that uses thermal stimulation method, the ratio of the amount of energy that can be produced from combustion of the total produced gas to the amount of input energy to the system during dissociation is introduced as the energy efficiency ratio (Equation 44) [132]:

$$\eta_r = \frac{V_{rp} Q_g}{Q_{rt}(2\pi r_7 h)}, \tag{44}$$

where $\eta_f$ is the energy efficiency ratio, and $Q_g$ is the heating value of the gas at STP conditions (J/m$^3$).

## Results and discussion

Pressure and temperature at the dissociation interface ($P_s$ and $T_s$) are coupled to the pressure and temperature at the wellbore (outer surface of the wellbore) due to their associated equations, mentioned earlier in the previous section. Thus, The dimensionless position and dimensionless velocity ($v_s / (4\alpha_{II} t)^{1/2}$) of the dissociation interface, both represented by $\beta$ (equation 35), which are connected to $P_s$ and $T_s$ (equation S27), are related to the pressure and temperature at the wellbore. The temperature at the outer surface of the wellbore is dependent on the temperature of the annulus section and time. The temperature inside the wellbore changes due to time and the temperature at the wellbore, which changes over time as the dissociation progresses. Table 1 shows the proposed properties and parameters for the base model according to previous studies [6, 70, 99, 101, 140-143].

As shown in Figure 3a, $\beta$ decreases at the beginning of the process because the thickness of Zone I increases over time, and the associated matrix in this zone absorbs a higher amount of transferred heat. Thus, dissociation is responsible for most of the energy consumption in the early stages of the process. But the energy consumption converges to a fixed value because the temperature at the wellbore surface is also converging slightly to that of inside of the well ($T_{fa}$). Figure 3b shows the location of the interface ($S$) during the process. The slope in the diagram is the highest at the beginning, but it decreases over time, which supports the observations from previous studies [24, 54, 132]. This behavior is in line with $\beta$, which decreases significantly at the beginning but slightly converges to a constant value and can be justified for the same reasons. Wang et al. [21] and Li et al. [144, 145], who experimentally investigated the hydrate dissociation upon thermal stimulation, also proved that at the beginning of the dissociation process, the rate of dissociation close to the wellbore is the highest. Our previous radial and flat analytical studies [75, 76] reported variable dissociation rates ($\beta$) due to the variable temperature at the wellbore, but it started from a lower value at the beginning of the process, then increased and converged to a higher value over dissociation. This different behavior could be due to the different heat source conditions employed in the present work. Actually, in our previous works [75, 76], the temperature of the heat source was assumed to remain constant, but in the present work, it increased over time. Furthermore, at the beginning, the dissociation rate is the highest, but the temperature increment of the heat source is not high enough to induce sufficient heat transfer to the dissociation front and raise the dissociation rate over time. When hot water is injected through the annulus, temperature distribution through this section, from which the heat transfer to the reservoir takes place, is higher, as shown in Figures 3c and 3d, inducing slightly higher values for $\beta$ and $S$ compared to those of the other operating scheme. The temperature difference becomes significant through the extraction part especially at the beginning of the process (Figures 3c and 3d). However, the associated differences decrease as the process continues to 100 days. This behavior was also reported by Beier et al. [101] who analytically and experimentally investigated the temperature distribution in the coaxial heat exchangers as ground source heat pumps.

The results presented in Figure 3 show another point related to the assumption of using average temperature along the wellbore in the calculations: a constant temperature along the wellbore surface causes a uniform vertical moving dissociation-interface. Hence, an inconstant temperature along the wellbore surface will cause a non-uniform dissociation interface, on which the water flow and the associated convection heat transfer would also impact. Thus, considering an inconstant temperature along the wellbore, instead of the average temperature, in the calculations requires taking into account the fluid flow in the reservoir, which lies out of scope of the present research, in order to achieve more accurate results.

Table 1. Parameters used for the base model.

| | |
|---|---|
| $r_1$, m | 0.043 |
| $r_2$, m | 0.05 |
| $r_3$, m | 0.07 |
| $r_4$, m | 0.077 |
| $r_5$, m | 0.092 |
| $r_6$, m | 0.099 |
| $r_7$, m | 0124 |
| Methane hydrate thickness, m | 80 |
| Thermal conductivity of inner tube (polyethylene tube), $k_p$, W/(m.K) | 0.4 |
| Thermal conductivity of cement, $k_c$, W/(m.K) | 0.933 |
| Thermal conductivity of gravel, $k_g$, W/(m.K) | 0.4 |
| Thermal conductivity of casing (steel), $k_s$, W/(m.K) | 43.3 |
| Thermal conductivity of water, $k_f$, W/(m.K) | 0.667 |
| Density of water, $\rho_f$ , kg/m$^3$ | 971.79 |
| Dynamic viscosity of water, $\mu_f$ , Pa.s | $3.54\times10^{-4}$ |
| Specific heat capacity of water, $C_f$, kJ/(kg.K) | 3.89 |
| Porosity, $\phi$ | 0.3 |
| Water flow rate, $V_f$ , L/s | 0.56 |
| Permeability, $k$, μm$^2$ | 1 |
| Thermal diffusivity of Zone I, $\alpha_I$ , μm$^2$/s | $2.89\times10^6$ |
| Thermal conductivity of Zone I, $k_I$ , W/(m.K) | 5.57 |
| Thermal diffusivity of Zone II, $\alpha_{II}$ , μm$^2$/s | $6.97\times10^5$ |
| Thermal conductivity of Zone II, $k_{II}$ , W/(m.K) | 2.73 |
| Hydrate density, $\rho_H$ , kg/m$^3$ | 913 |
| Heat of dissociation of hydrate, $Q_{Hd}$ , J/kg | $446.12\times10^3 - 132.638T_s$ |
| Gas heat capacity, $C_{pg}$ , J/(kg.K) | 8766 |
| Dynamic viscosity of gas, $\mu$ , Pa.s | $10^{-4}$ |
| Heating value of the gas at STP conditions, $Q_g$ , MJ/m$^3$ | 37.6 |
| Molecular mass of methane, *m*, g/mol | 16.04 |
| Mass ratio of the methane gas trapped inside the hydrate to the mass of hydrate, $F_{gH}$ | 0.1265 |
| Universal gas constant, R, J/(mol.K) | 8.314 |

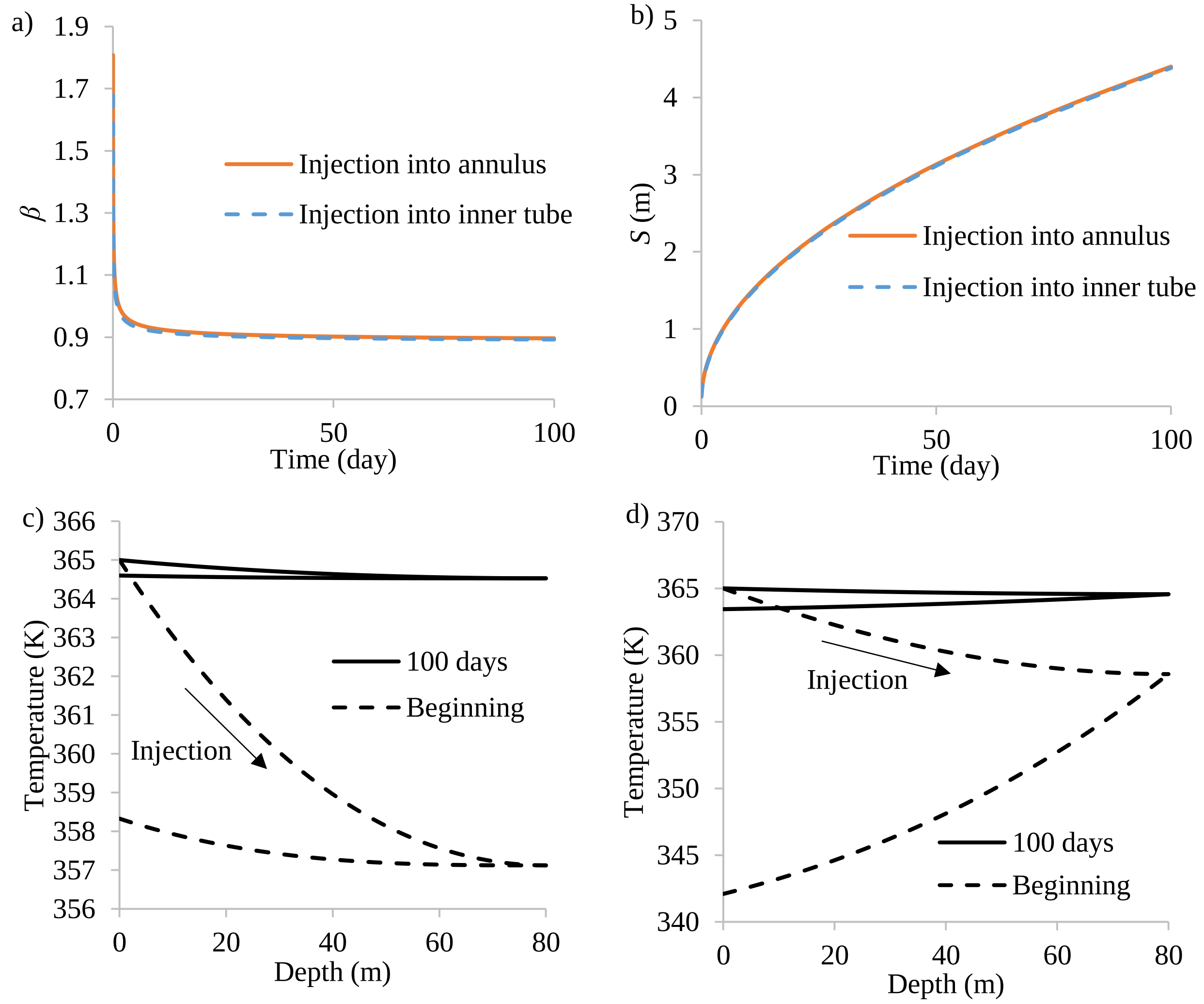

Figure 3. Variation of a) $\beta$ and b) interface location versus time for the two operation schemes. Temperature distribution in the wellbore versus depth when hot water is injected into c) inner tube and d) annulus. The conditions are: $T_0$=280 K and $T_i$=365 K, $P_i$=7.6 MPa, and a water flow rate of 0.56 L/s.

Figure 4a shows that decreasing the pressure at the wellbore and increasing the temperature of injected water decrease the value of $\beta$ after 100 days operation. Further information on how the injected water temperature affects dissociation is provided in the following results. The initial temperature of MH also has a direct effect on $\beta$ as shown in Figure 4b. These results are consistent with those of the previous work by Selim and Sloan [70] with the same trend, but $\beta$ is lower (approximately 48%) in the present model. The same behavior for $\beta$ against various initial pressures and temperatures was reported in our previous analytical studies [75, 76], but with higher values approximately 66% and 40% respectively in our previous flat [75] and radial [76] studies. These differences may be due to the following reasons: i) Selim and Sloan [70] considered a 1D flat heat source without wellbore thickness; and ii) all those works [70, 75, 76] assumed a very high and constant heat-source temperature.

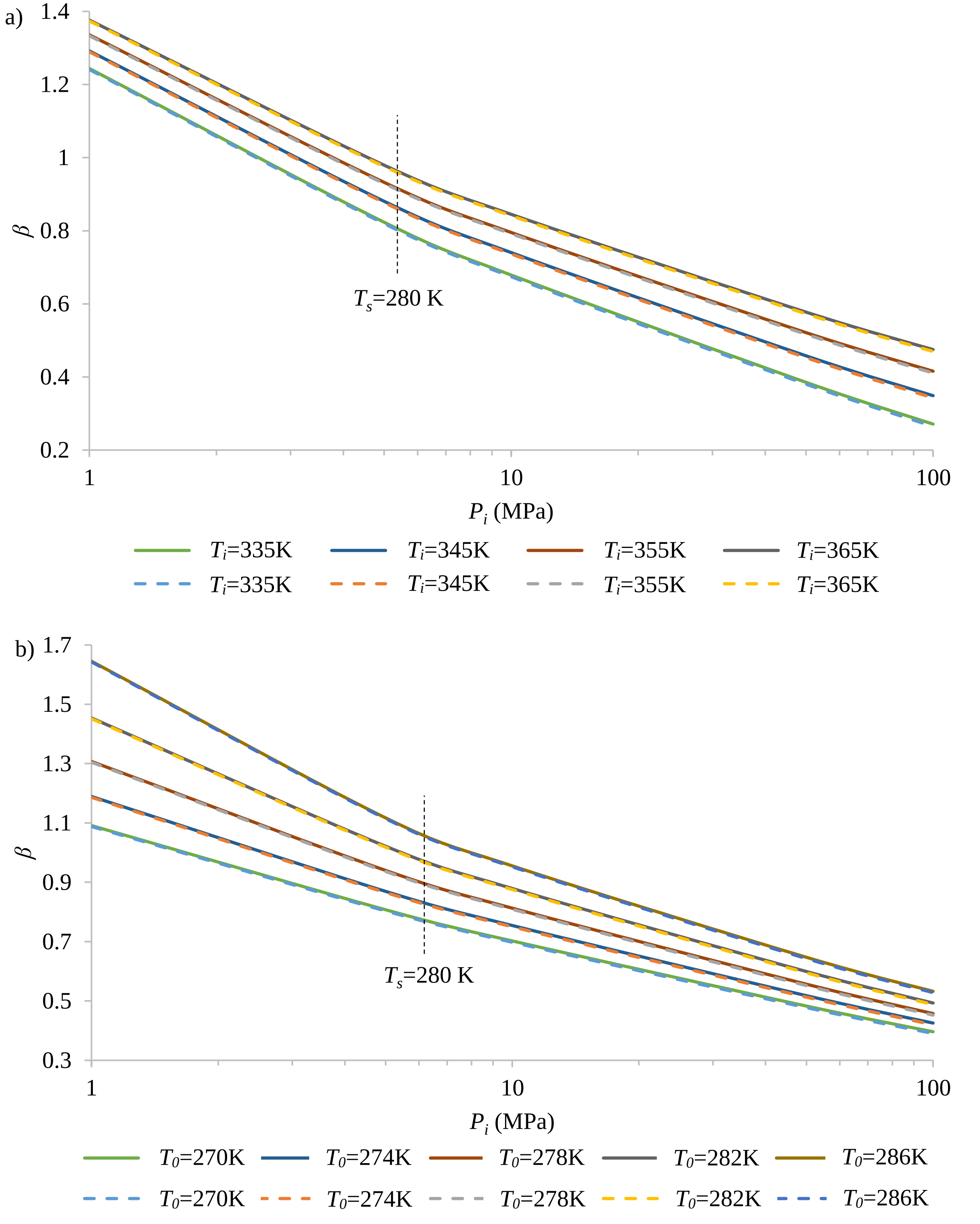

Figure 4. Dimensionless interface position at: a) $T_0$=280 K and various $T_i$ and $P_i$ values, and b) $T_i$=365 K and various $T_0$ and $P_i$ values. Dashed lines and solid lines respectively represent the cases with hot water injection into the inner tube and the annulus.

Furthermore, the locus at which the dissociation temperature is equal to 280 K (MH equilibrium temperature) is also shown on Figure 4. At the points on the left and right sides of this locus, the dissociation temperature decreases and increases, respectively. The dissociation temperature may depend on heat source pressure, whereas it is almost independent from the temperature of the heat source and MH. This result is in agreement with those of previous works [70, 75, 76], which also reported that at heat-source pressures lower than 6 MPa, $T_s$ may reduce to the freezing point of water, and the ice generation can halt dissociation. For the situations, in which $T_s$ is higher than the MH's temperature, some part of input heat from the heat source will be consumed to bring the temperature of MH zone close to the dissociation front up to $T_s$. If $T_s$ approaches $T_0$, all heat from the heat source is consumed for dissociation and no heat will be transferred to or from the MH zone near the dissociation interface, because the temperature of the hydrate zone remains constant. On the other hand, if $T_s$ falls below $T_0$, some part of the heat required for dissociation will be provided from the hydrate zone, causing in the temperature of the hydrate zone near the dissociation interface to drop.

Temperature and pressure distributions calculated for the two operation schemes are respectively depicted in Figures 5 and 6 for different time frames, with the following two boundary conditions (BCs): BC 1) $T_i = 340$ K, $P_i = 10$ MPa, and $T_0 = 280$ K, and BC 2) $T_i = 365$ K, $P_i = 7.6$ MPa, and $T_0 = 275$ K. It should be noted that the temperature of the injected water is below the saturation temperature of water in the present work. Black dashed lines in Figure 5 represent the temperature at the dissociation interface. As mentioned earlier, $T_s$ depends on the temperature at the wellbore and does not change significantly (Figure 5) because the temperature increase at the wellbore becomes smaller in longer time frames for the reasons stated earlier for the trend of $\beta$ during the process. Figures 5 and 6 along with Figure 4 show that the interface moves further if the inlet temperature is increased and the pressure at the wellbore is decreased, again reflecting reports by Selim and Sloan [70]. The interface pressure (Figure 6) is not constant and increases due to temperate changes at the wellbore surface. The pressure increment does not follow a constant slope and decreases as the temperature at the wellbore surface gets closer to that inside the well. Thus, it does not change significantly, resulting in convergence of the interface pressure. These results are in line with those reported by our previous works [75, 76], mainly because the temperature at the wellbore was not constant in those works due to the heat conduction through the wellbore structure. However, the increment of the temperature at the wellbore and interface pressure in the present work are lower than those of our previous works [75, 76] due to the reasons stated before for lower dissociation rates in the present work. In accordance with the present results, a study by Tsimpanogiannis and Lichtner [23], who built a semi-analytical model for hydrate dissociation, demonstrated that increasing the temperature of the wellbore raised the pressure at the interface. The temperature distribution differences between the two operation schemes is negligible, with only slightly higher temperature at the wellbore upon injection into the annulus, causing higher dissociation pressures and farther interface location from the wellbore.

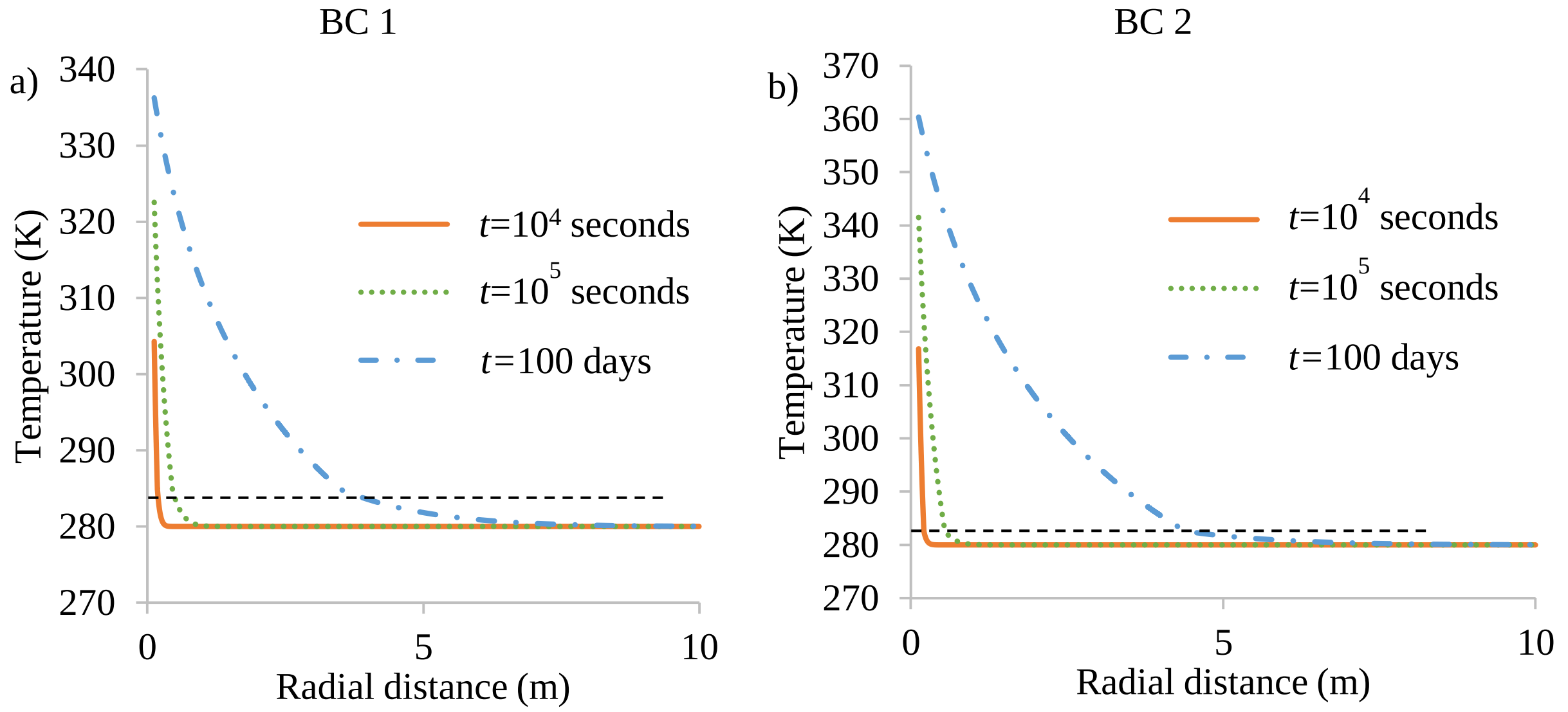

c)

d)

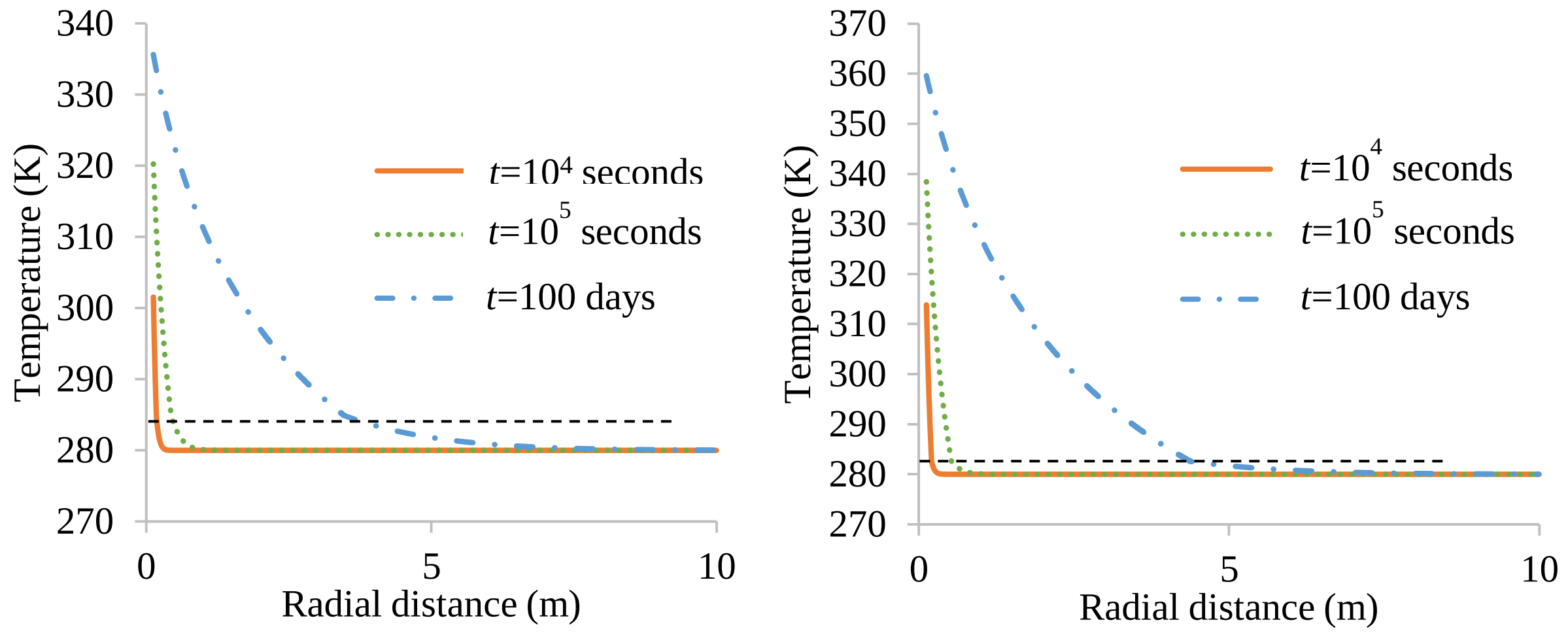

Figure 5. Temperature distribution at different time frames for two initial and boundary conditions (i.e., BC 1: a and c, BC 2: b and d). a) and b) are for hot water injection into annulus, and c) and d) are for hot water injection into the inner tube. The black dashed line specifies the temperature at the dissociation interface.

| BC 1 | BC 2 |
|---|---|

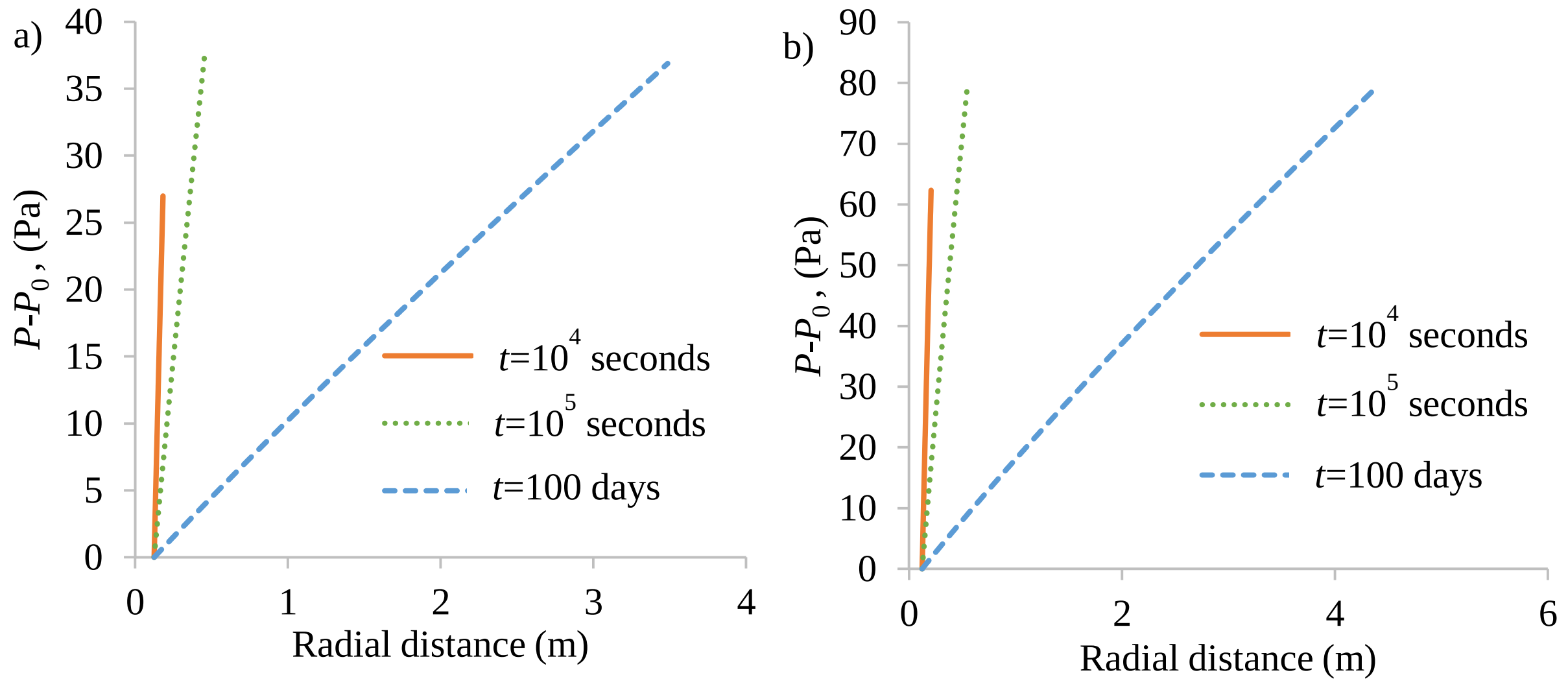

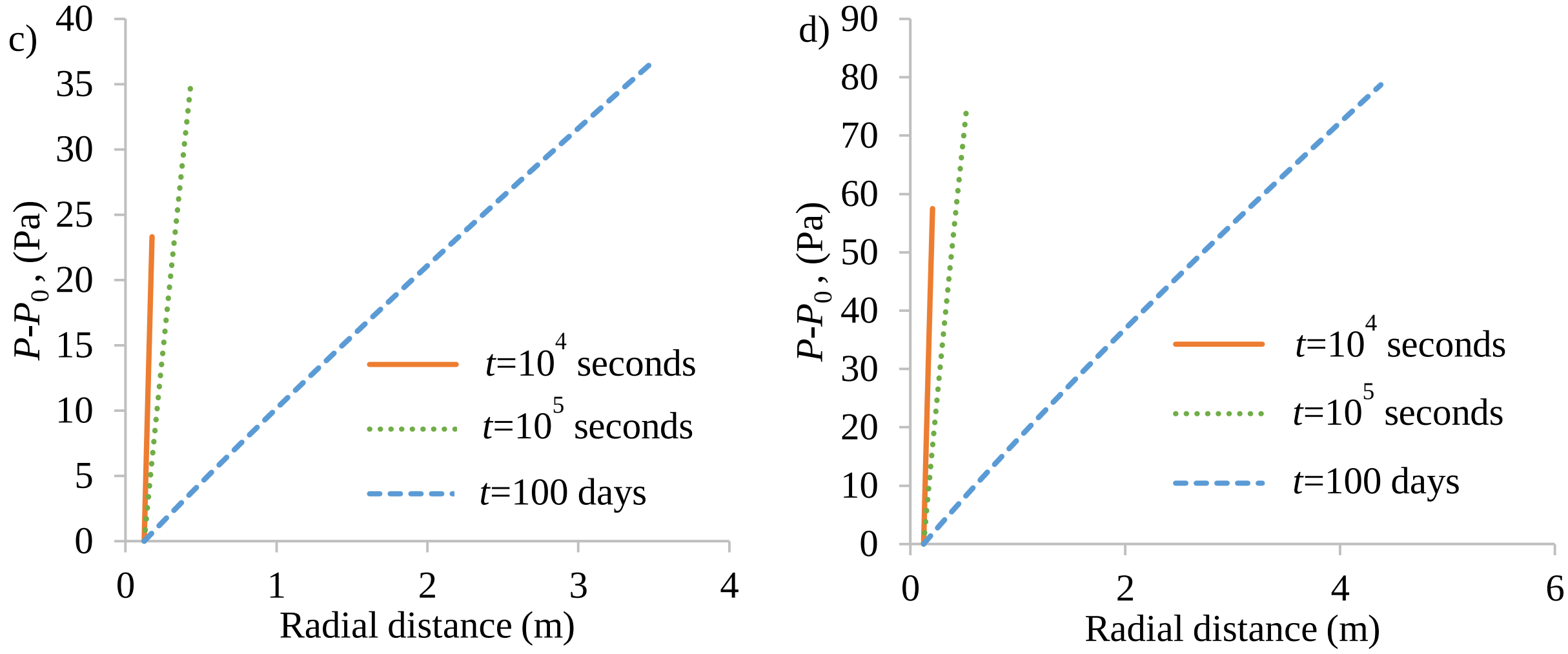

Figure 6. Pressure distribution in the dissociated zone at different time frames for two initial and boundary conditions (i.e., BC 1: a and c, BC 2: b and d). a) and b) are for hot water injection into annulus, and c) and d) are for hot water injection into the inner tube.

Figure S1 shows the total volume of produced gas ($m^3$) under STP conditions, and input heat ($MJ/m^2$) in two models with their specific operation schemes, considering the two initial and boundary conditions over 100 days. The produced gas volume is almost the same for both models, with slightly higher values for the case with hot water injection into the annulus, due to the direct heat transfer from heat source to reservoir. Beier et al. [101], who analytically investigated borehole heat exchangers and verified their results against experiments, reported the same heat transfer to the earth following injection into the inner tube or annulus. Holmberg et al. [99] performed and validated a set of numerical investigations on underground coaxial borehole heat exchangers and reported that different flow directions have almost the same efficiency. Figures S1a and S1b also show a negligible difference between the amount of heat transfer to the reservoir following injection into the inner tube or annulus. The produced gas and the input heat are higher under BC 2 than BC 1. The energy efficiency is higher in the model with BC 2 than that with BC 1 as shown in Figure 7, demonstrating that lowering the pressure at the wellbore and increasing the inlet temperature simultaneously result in more efficient gas production during dissociation. Although, our previous radial analytical model [76] showed the same trend for energy efficiency, higher energy efficiency by applying BC 1 was reported compared to that of BC 2. This behavior is due to the employing a coaxial wellbore heat source, which induced different temperature distribution along the wellbore and the associated heat transfer to the reservoir. Over time, the sediment matrix of Zone I becomes wider and absorbs a larger part of the input heat, causing reduction of the energy efficiency slope. This phenomenon causes the decrease in the dissociation rate (Figure 3a), which also causes the energy efficiency rate to drop over time (Figure 7) due to the continuous input heat increase from the wellbore to the reservoir (Figure S1). However, the energy efficiency does not drop because according to the associated formula for the energy efficiency (Equation 44) and Figure S1, the energy retrieved by the total produced gas consumption is higher than the input heat from the wellbore. Furthermore, Figure S1 shows that the input heat rate decreases over time, while the produced gas increases continuously with an almost constant rate, inducing energy efficiency increment during the process. Figure 7c shows a comparison of energy efficiency results of the previous experimental works and those of the case of hot water injection into the annulus in a shorter time frame.

Li et al. [146], who experimentally and numerically investigated gas production from MH using depressurization and thermal stimulation with hot water injection into horizontal wells, reported the net energy, which is the difference between the total retrieved energy by the produced-gas consumption and the total energy consumption in the dissociation, through MH dissociation is positive and increased continuously, but, the rate of net energy raise decreased. This report also implies a continuous raise and drop respectively in the energy efficiency and its associated rate because the total retrieved energy by produced gas consumption is higher than the total energy consumption in the dissociation. In an experimental work on gas production from MH using depressurization and electrical heating with vertical wells [84], the same results were also reported using the net energy calculation through the process. Fitzgerald et al. [147] through experiment on MH dissociation using a point heat source reported energy efficiency reduction from a high value over time. The present results are consistent with the experimental data of Li et al. [144, 145], who also employed huff and puff method in a 5.8 L cubic reactor. Li et al. [145] reported that the energy efficiency increases rapidly over 20 minutes after the beginning of the process, then starts to decrease. They also stated that the rapid raise in the energy efficiency early in the beginning of the process is caused by the fast hydrate dissociation rate close to the well, through which hot water is injected. Li et al. [144] reported that the energy efficiency raises until the end of the production stage, then drops until the end of the process. The calculated energy efficiency of the process was approximately 20.6 at the end of the process, and the produced gas had the similar trend as that of the present work. The present energy efficiency results show an approximately 39% and 83% mean difference respectively from those of Li et al. [144] and Li et al. [145] (Figure 7c), although, the present energy efficiency results tend to gradually converge to those reported by them [144, 145] as the dissociation progresses (Figure 7c). These relatively high differences are due to the experimental conditions, for example employing only huff and puff method [144] and direct hot water injection into the reservoir accompanied by depressurization in production stage [145]. Song et al. [132] by using this method reported the same trend for energy efficiency in their experimental work. The energy efficiency in their work was between 18 and 40. Comparison of gas production results with those of Wang et al. [24], who performed an experimental work employing huff and puff method, shows the similar trend with a maximum during the beginning of the process when hot water is injected to the reservoir. Their energy efficiency results also shows a good agreement with the present ones (Figure 7c) with an approximate mean difference of 30%, which is due to the huff and puff method employed along heat stimulation method by direct hot water injection into the reservoir in their work. Wang et al. [21] reported an energy efficiency of between 6 and 20 during their experimental investigations on MH dissociation upon hot water injection using different initial and boundary conditions. They stated that the hot water injection caused fast raise of the dissociation rate (gas production rate) during the first few minutes of the dissociation, inducing the maximum in the energy efficiency over the same time period. According to Figure 7c, their energy efficiency results differ approximately 39% from those of the present study. This could be due to the direct hot water injection into the reservoir and a small depressurization effect applied in their work in production stage. Additionally, the present energy efficiency tends to converge to their results [21] over time (Figure 7c).

The associated difference between the present results and those of the previous works could be due to some experimental conditions, such as direct hot water (with constant temperature) injection and circulation into the reservoir, which increases the heat loss and reduces the heat transfer to undissociated region (due to the blockage effect of water which will reduce the conduction heat transfer [128]), or employing other thermal stimulation methods (e.g., huff and puff) and their combination with depressurization. Furthermore, Figure 7c shows the experimental and analytical results for a short time period (a few hours), which is not close to real practical conditions, tend to converge steadily, and if the experiment continues to longer time frames they could follow the similar trend as that of the present work. It also should be noted that the wellbore structure, geometry, and performance mechanism (inside thermal and fluidic processes), which have been considered in the present study, have also noteworthy effects on those differences.

Selim and Sloan [70] in their analytical work reported an energy efficiency between 6.4-11.2, which remained constant during the process. Some of the assumptions in their work, which are mentioned earlier, are the reasons of the difference between the results of the present study from theirs.

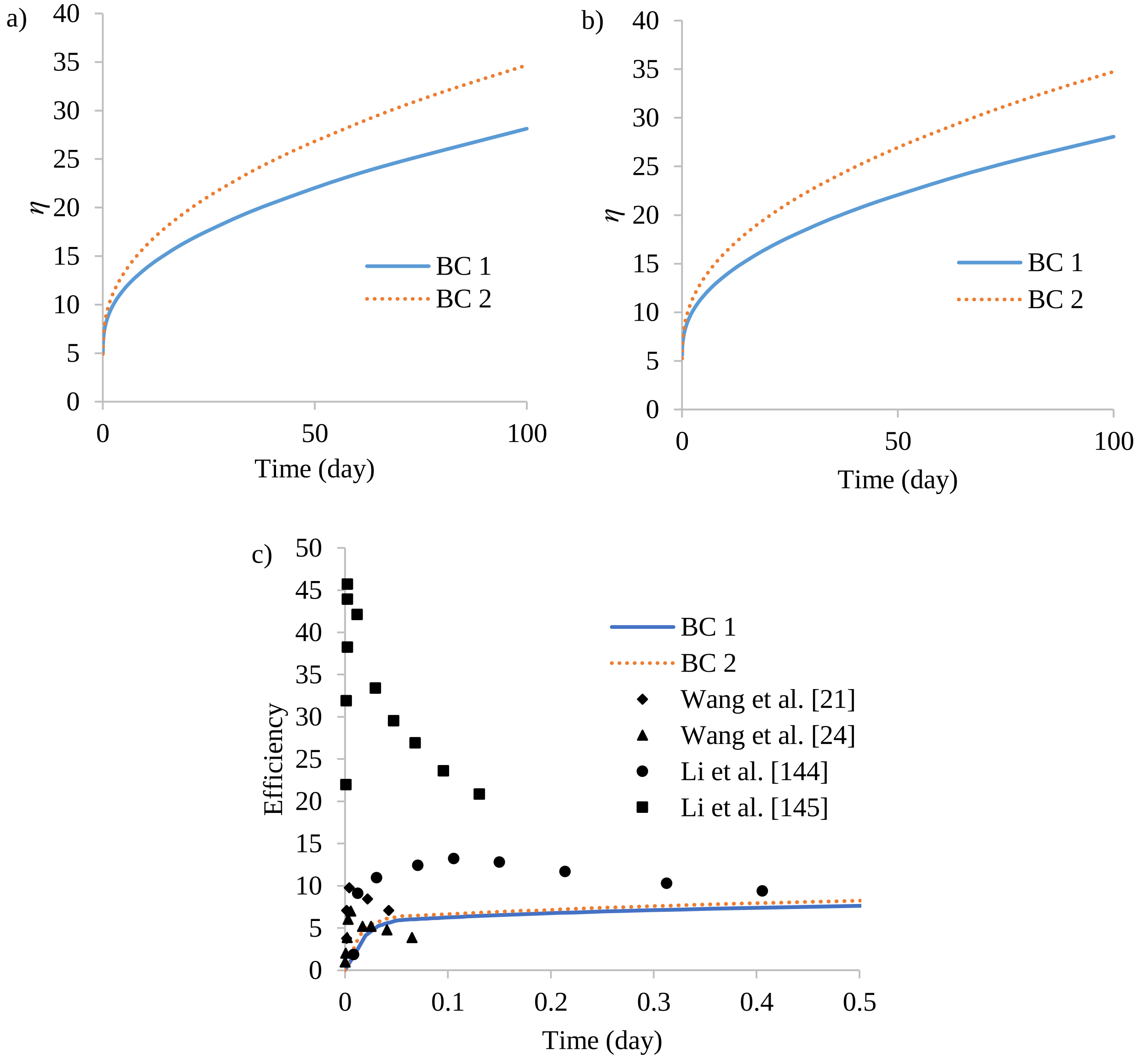

Figure 7. Energy efficiency during hydrate dissociation for the two BCs in the model with hot water injection into the a) annulus and b) inner tube after 100 days of process. c) energy efficiency of the case with hot water injection into the annulus in a shorter time frame compared to the results of the previous experimental works (adopted from: Applied Energy, Vol 110, Yi Wang, Xiao-Sen Li, Gang Li, Yu Zhang, Bo Li, Zhao-Yang Chen, Experimental investigation into methane hydrate production during three-dimensional thermal stimulation with five-spot well system, 90-97, Copyright (2013), with permission from Elsevier; Fuel, Vol. 117 (A), Yi Wang, Xiao-Sen Li, Gang Li, Ning-Sheng Huang, Jing-Chun Feng, Experimental study on the hydrate dissociation in porous media by five-spot thermal huff and puff method, 688-696, Copyright (2014), with permission from Elsevier; Energy, Vol. 64, Gang Li, Xiao-Sen Li, Bo Li, Yi Wang, Methane hydrate dissociation using inverted five-spot water flooding method in cubic hydrate simulator, 298-306, Copyright (2014), with permission from Elsevier; Energy & Fuels, Vol. 26 (2), Xiao-Sen Li, Yi Wang, Gang Li, Yu Zhang, Experimental Investigations into Gas Production Behaviors from Methane Hydrate with Different Methods in a Cubic Hydrate Simulator, 1124-1134, Copyright (2011), with permission from American Chemical Society.

A parametric study has been designed based on the parameters listed in Table 2, to investigate the effect of various characteristics of reservoir and wellbore on the dissociation process. From Figures 8 and 9: i) the higher the thermal diffusivities and conductivities of Zone I, the higher the dissociation rate and gas production will be; ii) the lower the thermal diffusivities and conductivities of Zone II, the higher the dissociation rate will be, on the other hand, higher thermal diffusivities and lower thermal conductivities induce more gas production; iii) the dissociation rate and gas production have almost no connection to the permeability and gas viscosity; and iv) the porosity of the media has a direct relation to the dissociation rate, in contrast, it has an inverse relation with the gas production.

Figure S2 in the supplementary file displays the input heat from the wellbore to the reservoir during the parametric study of the reservoir's characteristics. Higher thermal conductivity in Zone I induces significantly higher amounts of input heat from the reservoir (Figure S2a). Higher thermal diffusivity (lower heat capacity while the density is constant) induces less heat storage in the media (lower input heat), which in turn, causes more heat transfer to the dissociation front. Thus, increasing the thermal diffusivity and thermal conductivity of Zone I increases the dissociation rate (Figure 8a) and gas production level (Figure S2a). Lower thermal diffusivity of Zone II increases the dissociation rate (Figure 8b) due to greater storage of the transferred heat to this zone from the dissociation interface, which in the end, will be released and consumed for dissociation. Increasing the thermal conductivity of Zone II causes faster heat transfer to this zone from the moving interface, reducing the dissociation rate and gas production. Increasing the thermal diffusivity of Zone II reduces heat storage in this zone, increasing heat consumption for dissociation and raising gas production (Figure S2b). Figure S2b shows that decreasing the thermal diffusivity and increasing the thermal conductivity of Zone II, resulting in more input heat for the same reason stated for input heat increment induced by the same change in thermal conductivity and diffusivity in Zone I (Figure S2a). However, the amount of input heat increment is much lower for the case of Zone II than that of Zone I due to the direct contact of Zone I with the wellbore. Conduction heat transfer and input heat from the wellbore decrease for higher porosities (Figure S2c), ultimately causing a reduction in the dissociation rate (Figures 8c and 8d). Furthermore, higher porosity increases the amount of MH trapped in the pores, eventually increasing the amount of gas produced (Figures S2c and S2d).

Figures 9a and 9b clearly show the direct relation between energy efficiency and thermal diffusivity in Zones I and II, caused by the rise in gas production and drop in input heat induced by increasing the thermal diffusivity. On the other hand, the energy efficiency has an inverse relation with the thermal conductivity of Zones I and II, perhaps due to: i) higher input heat induced by the higher thermal conductivities of Zone I (Figure S2a); and ii) the lower gas production caused by the higher thermal conductivities of Zone II (Figure S2b). The curves shown in Figures 8c, 8d, 9c, and 9d reveal that different permeability and gas viscosities have almost the same results and overlapped each other with negligible difference. Thus, there are only two curves representing the results for the two operating schemes in the figures, which are mentioned above. There is also no difference between the results obtained from different permeability and different gas viscosities. This also reveals that reservoir permeability and gas viscosity have no impact on the dissociation. The results shown in Figures 8 and S2 are in good agreement with those of our previous radial study [76], but, we reported that increasing the porosity of the media increased the energy efficiency of the process. This discrepancy could be related to the differences between the models as stated previously, and we believe that the conditions applied in the present work make the results closer to the real conditions and more reliable. It should also be noted that our previous flat analytical study [75] revealed the same dissociation rate as those shown in Figure 8.

Selim and Sloan [70] performed a similar parametric study and reported similar results for the rate of dissociation (Figure 8). Zhao et al. [126] mathematically showed that increasing the thermal conductivity had a direct positive effect on a dissociation process based on thermal stimulation. They also reported that almost no change occurred in the dissociation by changing the relative permeability of water and gas, due to the negligible impact that the convection heat transfer of water and gas has on the process. Zhao et al. [127] in another numerical work showed that increasing the sediments' thermal conductivity caused a higher gas generation rate at the beginning of dissociation employing depressurization. Both of their works were verified by the experimental data of Masuda's work [49]. Tsimpanogiannis and Lichtner [23], who performed a similar parametric study on MH dissociation upon thermal stimulation, showed that the higher the thermal conductivity of the porous media, the higher the MH dissociation will be. Moridis et al. [54] numerically showed that a higher initial formation temperature, well temperature, and formation thermal conductivity increased the amount of gas production at the Mallik site. It is also shown in Figures 8 and S2 (the dissociation rate ($\beta$) has a direct relation to the amount of produced gas). It should be noted that some of different working conditions, such as direct hot water circulation into the reservoir, the duration of experiments, and model parameters (i.e., hydrate saturation) may have caused the differences between the results of the experimental works and those of the present work.

Li et al. [97], who numerically investigated gas production from MH deposits of Qilian Mountain, Qinghai province, upon thermal stimulation reported that changing the intrinsic permeability of reservoir from 0.001 $\mu m^2$ to 0.01 $\mu m^2$ improved the cumulative amount of gas production about 13%. Moridis et al. [54] numerically performed sensitivity analysis on the impact of reservoir effective permeability on MH dissociation upon thermal stimulation from reservoirs located in Malik site. They reported that changing the permeability from 0.02 $\mu m^2$ to 0.1 $\mu m^2$ has no specific impact on gas production. Li et al. [96] through their numerical investigation on gas production from hydrate-bearing layer, located in Shenhu Area, China, by thermal stimulation method, performed a sensitivity analysis by employing 0.0001 $\mu m^2$, 0.00075 $\mu m^2$, 0.04 $\mu m^2$, and 0.075 $\mu m^2$ reservoir intrinsic permeability. Their results showed that increasing reservoir permeability improved volumetric rate of methane release. However, the raise in volumetric rate of methane release became smaller in higher reservoir permeability. For example, the volumetric rate of methane release resulted from 0.075 $\mu m^2$ reservoir intrinsic permeability converged to the associated value resulted by considering 0.04 μm$^2$ reservoir permeability over time. Furthermore, the volumetric rate of methane release raised about 33% by changing the permeability from 0.04 μm$^2$ to 0.075, while it changed about 83% by changing the permeability from 0.00075 μm$^2$, 0.04 μm$^2$. Thus, it can be concluded that the dissociation process becomes insensitive to high intrinsic permeability of reservoir. It should also mentioned that the water produced from the dissociation is assumed to remain motionless, and only gas fluid flow in radial direction is considered. Hence, the permeability affects the dissociation only through gas flow in the reservoir. Furthermore, Selim and Sloan [70] in their analytical investigation on MH dissociation upon thermal stimulation reported that the convection heat transfer by gas flow decreased the dissociation rate about 6% meaning that it does not have a significant impact on the process.

Table 2. Range of parameters assumed in the parametric study.

| | |
|---|---|
| Reservoir parameters | |
| Porosity, $\phi$ | 0.1 to 0.5 |
| Permeability, $k$, μm$^2$ | 0.1 to 5 |
| Thermal diffusivity of Zone I, $\alpha_I$ , μm$^2$/s | $1\times10^6$ to $5\times10^6$ |
| Thermal conductivity of Zone I, $k_I$ , W/(m.K) | 3 to 7 |
| Thermal diffusivity of Zone II, $\alpha_{II}$ , μm$^2$/s | $4\times10^5$ to $8\times10^5$ |
| Thermal conductivity of Zone II, $k_{II}$ , W/(m.K) | 1 to 5 |
| Gas viscosity, $\mu$ , Pa.s | $10^{-4}$ to $10^{-6}$ |
| Wellbore parameters | |
| Water flow rate, $V_f$ , m$^3$/s | 0.0004 to 0.006 |
| Inlet temperature, $T_i$, K | 330 to 365 |
| Thickness of MH reservoir, $h$, m | 15 to 80 |
| Outer radius of the wellbore, $r_7$, m | 0.114 to 0.174 |
| Only the annulus radius, $r_7$, m | 0.114 to 0.164 |
| Only the inner tube radius, $r_2$, m | 0.013 to 0.053 |

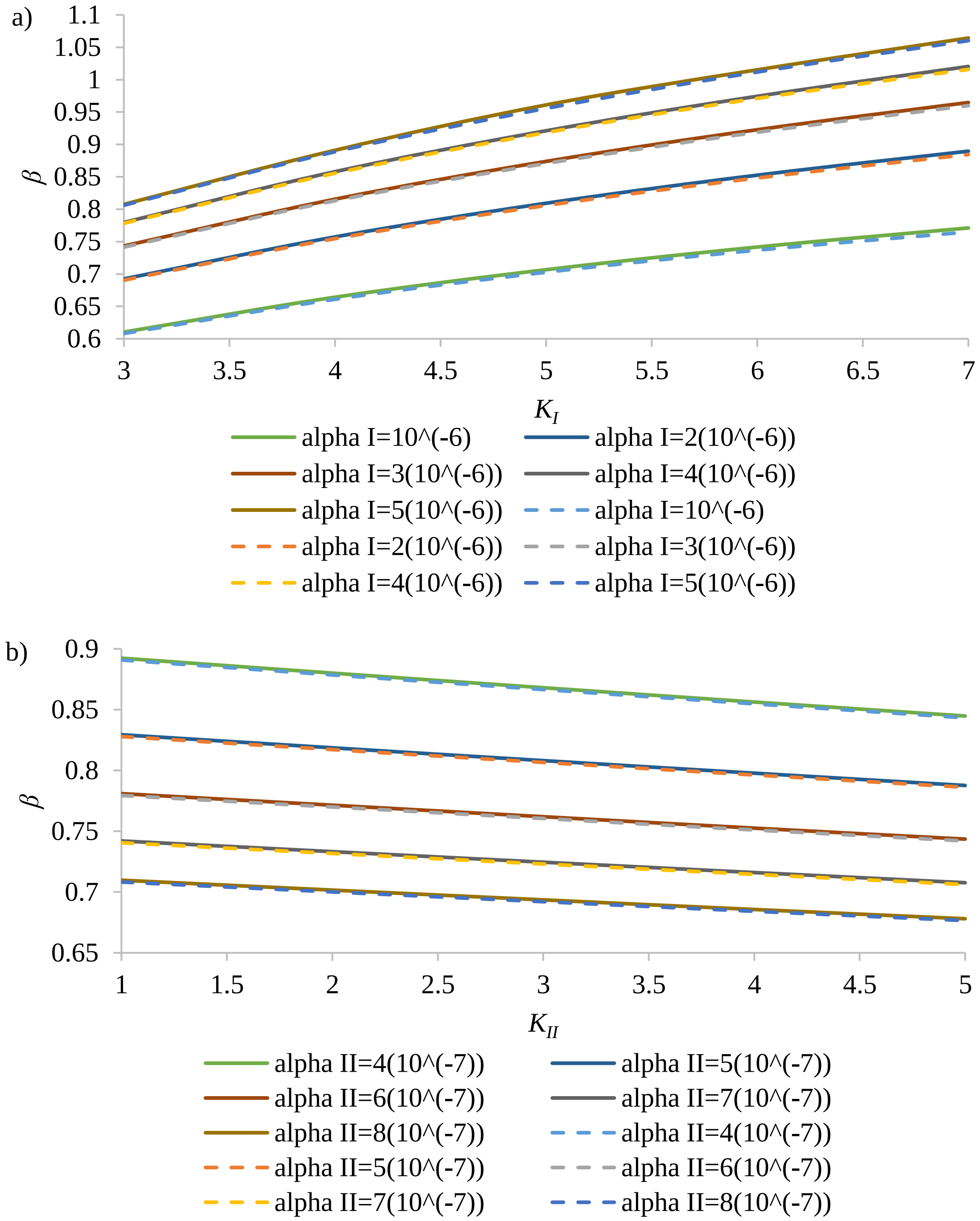

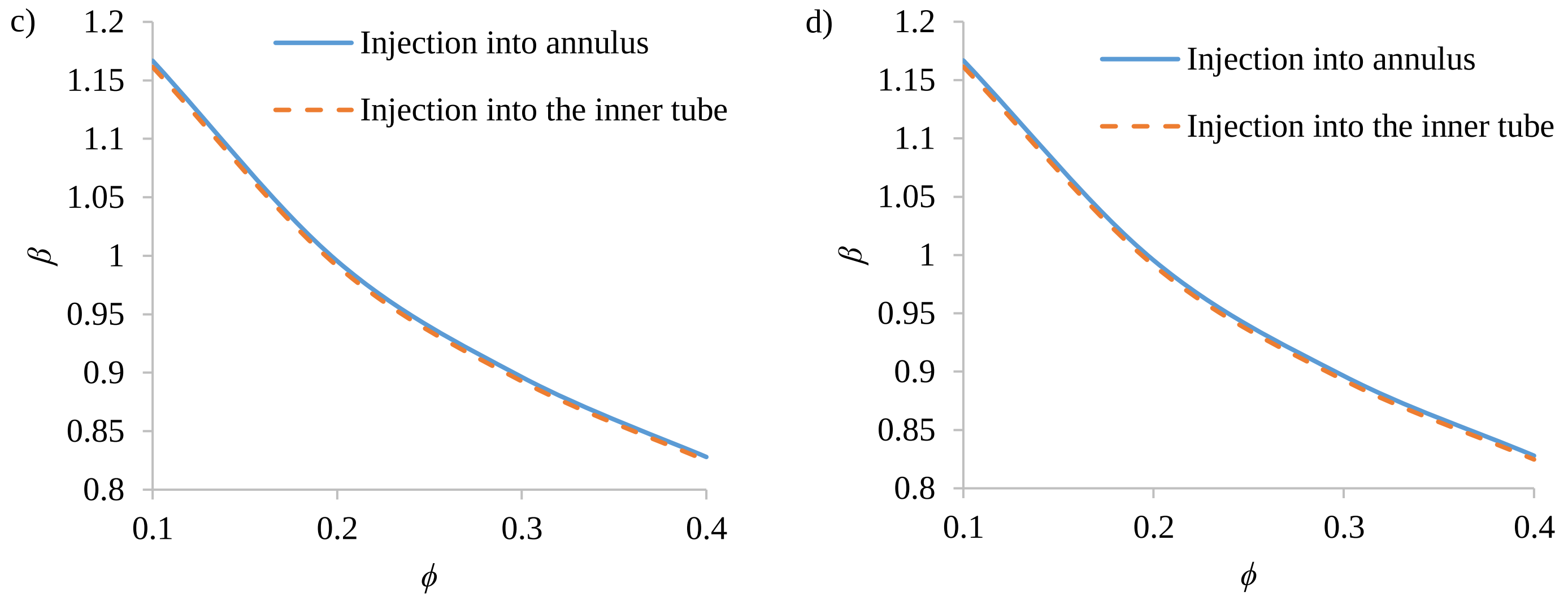

Figure 8. The effect of various parameters on the interface movement after 100 days dissociation considering both operating schemes: a) thermal diffusivity and thermal conductivity of Zone I, b) thermal diffusivity and thermal conductivity of Zone II, c) porosity with various permeabilities, and d) porosity with various gas viscosities.

Solid lines and dashed lines respectively represent the operating schemes of hot water injection into annulus and into the inner tube.

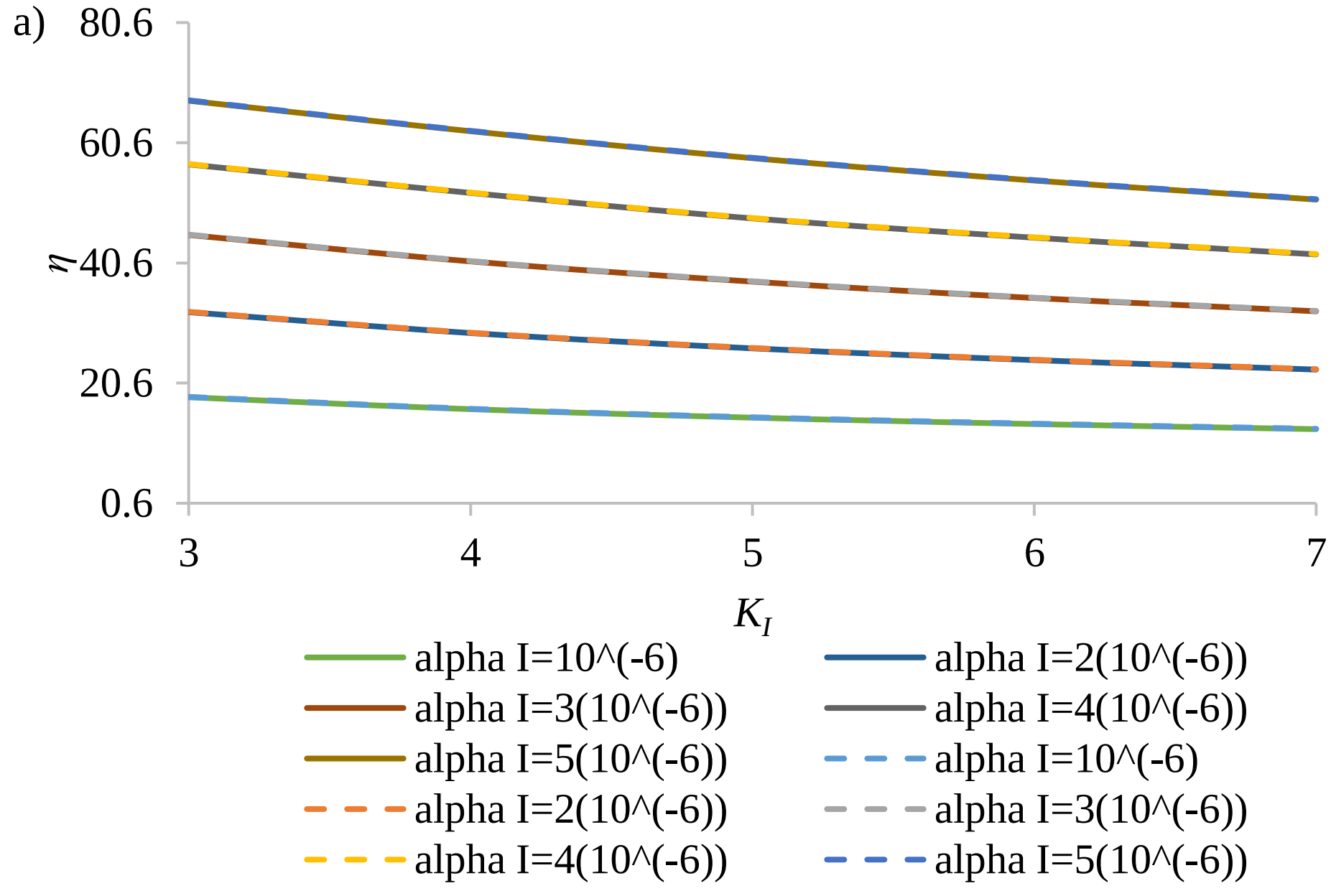

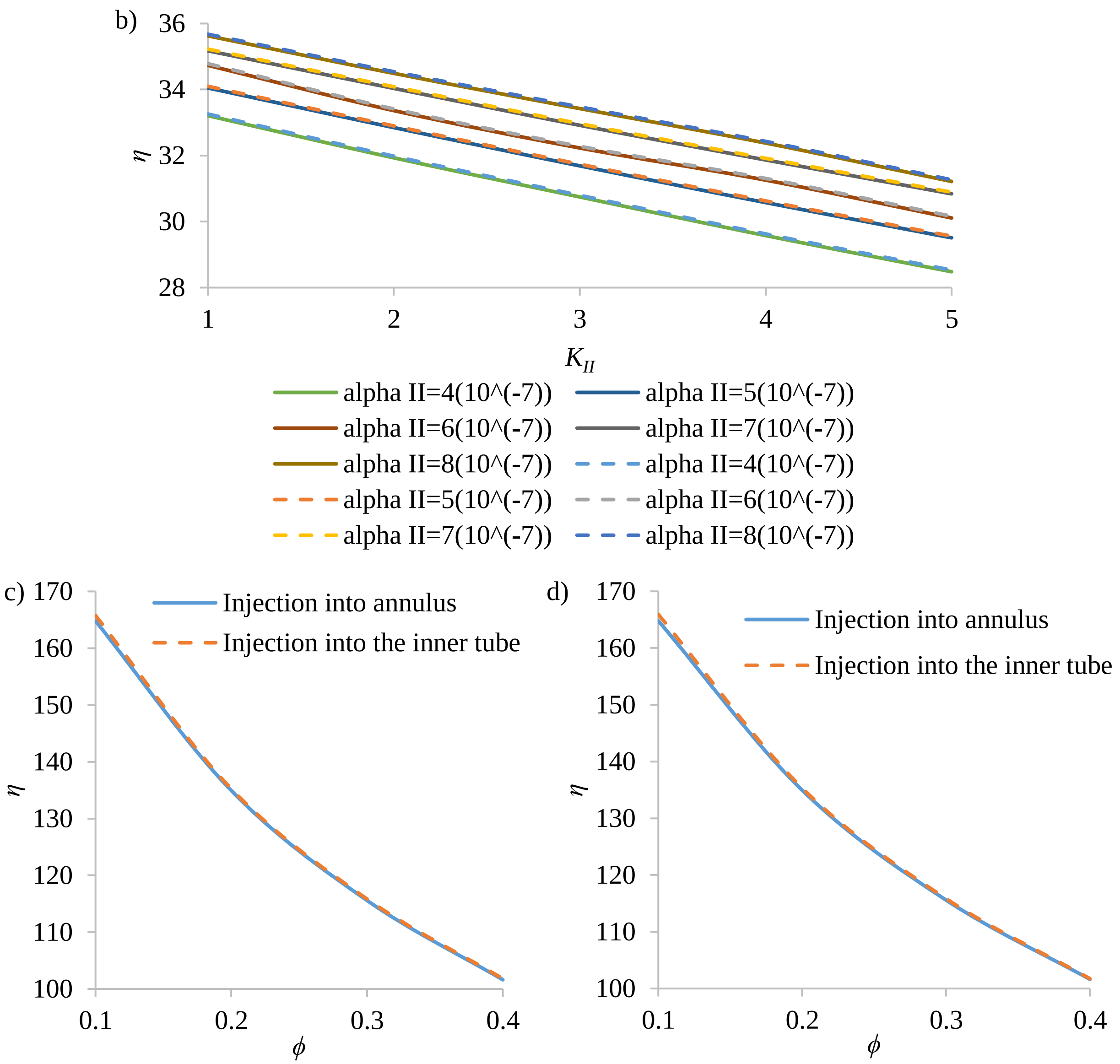

Figure 9. Energy efficiency after 100 days dissociation considering both heat sources and various parameters: a) thermal diffusivity and thermal conductivity of Zone I, b) thermal diffusivity and thermal conductivity of Zone II, c) porosity with various permeabilities, and d) porosity with various gas viscosities.

Solid lines and dashed lines respectively represent the operating schemes of hot water injection into annulus and into the inner tube.

Figures 10, 11, and S3 show the results of the parametric study examining how the wellbore parameters (Table 2) affect dissociation. This parametric study has not been performed in any of the previous works on MH dissociation, and the associated results highlight the following points: i) a higher wellbore radius (annulus and inner tube with the same amount) increases the dissociation rate (Figure 10a) and gas production level (Figure S3a) because the contact surface enlargement affects the process significantly; ii) the dissociation rate decreases in reservoirs with larger thickness (Figure 10b), whilst, the gas production increases due to the contact surface enlargement (Figure S3b); iii) enlarging the annulus radius, while the inner tube's radius is fixed, has a direct relation on the dissociation rate (Figure 10c) and produced gas (Figure S3c) due to the contact surface enlargement; iv) higher inner tube's radii, while keeping the annulus's radius fixed, has almost no effect on the dissociation process. Raymond et al. [148] through analytical investigations, revealed that increasing the inner tube's radius has little effect on the thermal resistance of coaxial borehole heat exchangers; v) higher inlet temperatures has a direct and noteworthy impact on both the dissociation rate (Figure 10e) and produced gas (Figure S3e); vi) the dissociation rate and gas production increase initially in response to a transition to higher water flow rates, but they tend to remain almost unchanged after a certain water flow rate.

Increasing the wellbore radius reduces the velocity of the hot water flow inside the wellbore, which in turn reduces the mean temperature inside the wellbore. As a consequence, the input heat from the wellbore to the reservoir drops as well (Figure S3a). The energy efficiency of the process decreases as the wellbore radius increases (Figure 11a) because the ratio of produced gas to the input heat also decreases.

A higher reservoir thickness decreases the input heat (Figure S3b), due to the reduction in the mean temperature inside the wellbore induced by longer wellbores. Holmberg et al. [99] also reported that the deeper coaxial borehole heat exchangers are not good at heat transfer into the reservoir, but are useful for heat extraction from the reservoir. Raymond et al. [148] analytically proved that longer coaxial boreholes have higher thermal resistances. On the other hand, increasing the reservoir's thickness boosts energy efficiency slightly (Figure 11b) because there is a greater rise in gas production than in input heat. The differences in results between injection into the inner tube and injection into the annulus increases with an increase in the reservoir thickness. Actually, in longer wellbores, hot water injection into the inner tube makes the drop in the flow mean temperature in the annulus become more pronounced because the flow enters the annulus with a lower temperature, while in the other case, the hot water inlet is the annulus, which is in direct contact with the reservoir.

The input heat decreases by increasing only the annulus radius (Figure S3c) due to the same reason stated previously for the effect of the wellbore radius on the input heat. This observation also accords with another analytical study on ground coaxial heat exchangers conducted by Raymond et al. [148]. They reported that increasing only the annulus radius reduced the wellbore thermal resistance. In contrast to the higher produced gas induced by bigger annulus radius, the energy efficiency decreases (Figure 11c). Bigger annulus radii while keeping the inner tube's radius fixed will reduce the flow rate in the annulus significantly, reducing the mean temperature in annulus, and it causes more input heat reduction compared to the case of increasing the radius of both annulus and inner tube together.

The input heat and the energy efficiency increase by increasing the inlet temperature (Figures S3e and 11e). The difference of energy efficiency and input heat results between the two operating schemes tends to get slightly bigger as the inlet temperature increases. Increasing the inlet temperature in the case of hot water injection into annulus directly affects the input heat, but in the other case it causes higher mean temperature in the inner tube, and the associated temperature increase in annulus is not significant.

The input heat increases by increasing the flow rate (Figure S3f), but it converges to almost a fixed value. It can be concluded that the mean temperature inside the wellbore initially increases but increasing the flow rate more than 0.0016 $m^3/s$ almost has no effect on the mean temperature inside the wellbore and the associated input heat to the reservoir. Zanchini et al. [149], who carried out numerical analysis on the performance of coaxial borehole heat exchangers, showed that the heat transfer capacity of the wellbore would be increased by increasing the flow rate of the wellbore. This is also in line with the results of an analytical work performed by Raymond et al. [148], who showed that the higher flow rates decreased the thermal resistance of coaxial ground heat exchangers. Despite the produced gas increment due to the higher flow rates, energy efficiency of the process reduces (Figure 11f) because the input heat increment is slightly higher than that of the produced gas.

The amount of produced gas and input heat are always higher for the case of injection into annulus than those of the other operating scheme, but, the resulted energy efficiency is always slightly higher for the case of injection into the inner tube except for the cases of studying different flow rates. In fact, by increasing the flow rate in the annulus or in the inner tube, the flow rate would be increased through the wellbore by the same ratio, and it does not depend on hot water inlet.

The abovementioned results reveal the significant role played by the wellbore structure and its associated working conditions (i.e., hot water circulation in the wellbore and heat transfer process to the reservoir) in MH dissociation upon wellbore heating by hot water circulation. This is also an important aspect in field-test investigations, which deal with real equipment along with real conditions. For example, one vertical well was designed and employed to induce depressurization and thermal stimulation by hot air and steam injection for gas production from hydrate deposits in the Qilian Mountain permafrost [150]. They put several well screens in various locations of the well casing. In another gas production field-test in marine hydrate deposits of Nankai Trough, specific wellbore designs were employed for specific purposes [15, 151]. For instance, the production well contains a pump and a heater [15]. Hence, wellbores become more complex by taking into account the field tests due to various parameters, such as geology of the region and the proposed dissociation method, requiring more in-depth studies on wellbores in advance to the field operation in order to achieve a better prediction of the outcomes and optimization of the process.

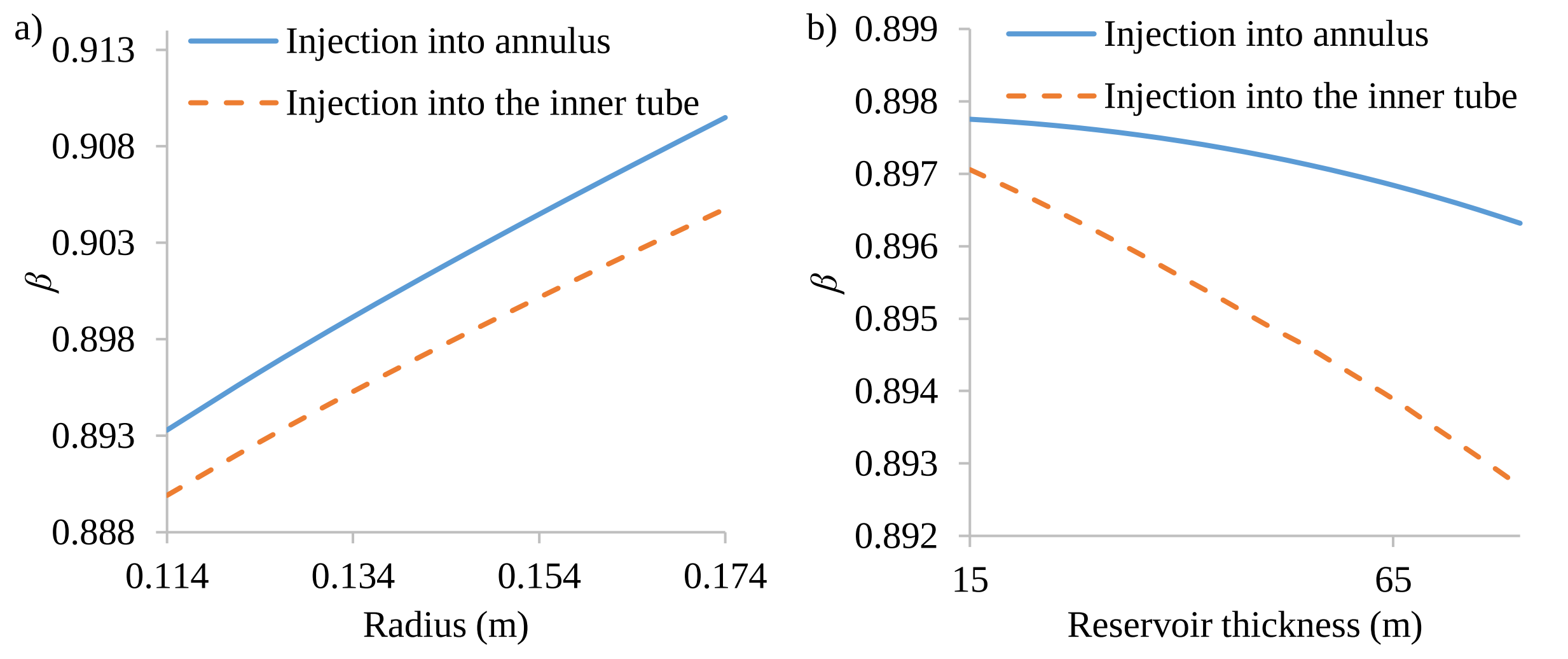

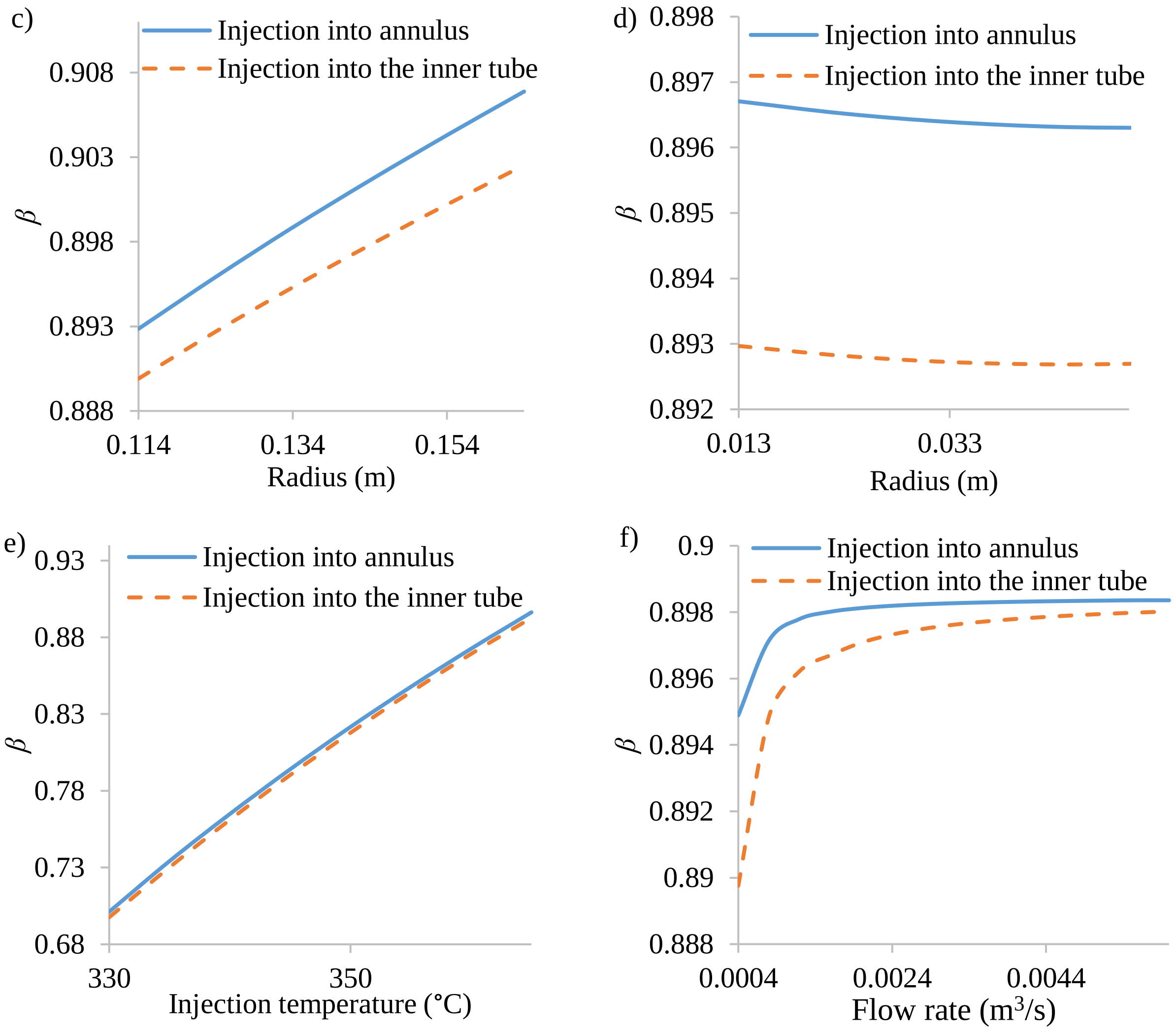

Figure 10. The effect of various parameters on the interface movement after 100 days dissociation considering both operating schemes: a) the wellbore radius, b) the reservoir thickness, c) the annulus radius, d) the inner tube radius, e) the injection temperature, and f) the flow rate.

Dashed lines and solid lines are respectively representative of the operating schemes of hot water injection into annulus and into the inner tube.

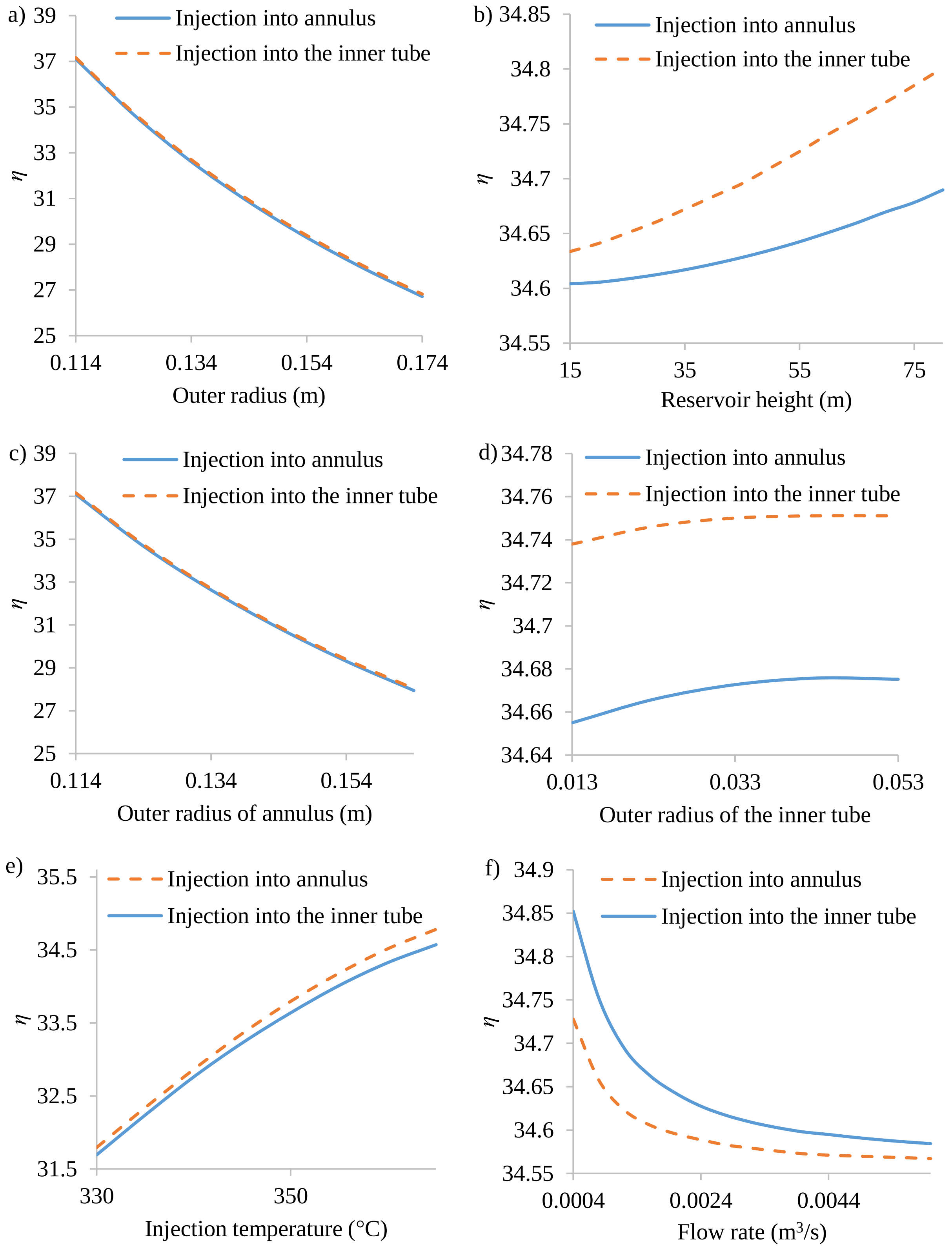
a)
Injection into annulus
Injection into the inner tube
η
Outer radius (m)
b)
Injection into annulus
Injection into the inner tube
η
Reservoir height (m)
c)
Injection into annulus
Injection into the inner tube
η
Outer radius of annulus (m)
d)
Injection into annulus
Injection into the inner tube
η
Outer radius of the inner tube
e)
Injection into annulus
Injection into the inner tube
η
Injection temperature (°C)
f)
Injection into annulus
Injection into the inner tube
η
Flow rate (m3/s)

Figure 11. The effect of various parameters on the energy efficiency after 100 days dissociation considering both operating schemes: a) the wellbore radius, b) the reservoir thickness, c) the annulus radius, d) the inner tube radius, e) the injection temperature, f) the flow rate.

Dashed lines and solid lines respectively represent the operating schemes of hot water injection into annulus and into the inner tube.

## Conclusions

The present study was designed to determine different aspects of MH dissociation upon thermal stimulation employing a coaxial wellbore structure as the heat source. For the first time, radial analytical models have been developed to couple operational conditions of the heat source and the associated MH dissociation in the reservoir. Heat and mass transfer in both wellbore and reservoir as well as the convection heat transfer in the wellbore are taken into account. Two operating schemes for wellbore heating are considered: i) hot water supply into the inner tube; and ii) hot water supply into the annulus section. The effects of various parameters of wellbore and reservoir as well as different boundary conditions on dissociation are also evaluated. Two main factors are considered to assess the production process performance: i) dissociation rate ($\beta$), and ii) energy efficiency of the process ($\eta$). Taken together, the most important findings to emerge from the results are as follows:

- The dissociation process has direct relation to the temperature of the reservoir and the inlet water, but it has an inverse relation to the wellbore pressure.
- The different operating schemes have approximately the same outcome with negligible differences due to the model conditions.
- Increasing the inlet water temperature while decreasing wellbore's pressure increases the dissociation rate and the produced gas. It also increases the energy efficiency despite some of the previous works' reports. This could be due to the inconstant wellbore temperature in the present work. Hence, more information about the heat source operating conditions can help increasing the accuracy on this matter.
- Different wellbore radii, water inlet temperatures and flow rates along with various properties, such as thermal diffusivities, porosities, thermal conductivities, and reservoir thicknesses have significant effects on the process. On the other hand, some of the reservoir and wellbore parameters, such as inner tube radius, reservoirs permeability, and gas viscosity, have almost no impact on the process. Thus, a thorough investigation on the reservoir properties and applicable heat source characteristics is necessary before performing a field work.

- Temperature at the wellbore surface changes over time due to the different temperatures inside the wellbore induced by convection and conduction heat transfer of the hot water flow inside the wellbore. It is also affected by the heat conduction from wellbore to reservoir, which is dependent on the wellbore outer structure.
- Temperature at the wellbore surface affects both the pressure and temperature at the dissociation front.
- The results of this study are in good agreement with those of the previous experimental and numerical studies. It also validates the assumptions made during the model development.

The findings of the present study contribute in several ways to our understanding of MH dissociation upon wellbore heating method and provide a basis for future investigations. The presented analytical approach takes into account: i) coupling the heat transfer process inside the coaxial wellbore and the associated thermal response in the reservoir which governs the dissociation and methane production; ii) heat transfer through the wellbore structure, consisting of multiple layers with different thermal properties; and iii) various wellbore and reservoir parameters. Some discrepancies exist between the results of the present work and those of our previous works [75, 76], such as temperature and pressure distributions inside the reservoir, temperature at the wellbore, energy efficiency of the process, and the results of parametric study. These differences could be related to the different wellbore characteristics (e.g., a coaxial wellbore heat-source and a constant-temperature heat source) and model geometries (e.g., radial and flat models) and conditions employed in the present work, which make this study more reliable, comprehensive, and closer to real practical conditions.

**Acknowledgement**

Financial support for this work provided by Natural Sciences and Engineering Research Council of Canada (NSERC)

**Nomenclature**

| | | | |
|---|---|---|---|
| $A$ | Dimensionless variable in equation 37 | $q_i$ | Heat transfer rate from/to the inner tube to/from the annulus, (MJ/(m$^2$.s)) |
| $A_a$ | Dimensionless constant in equation 24 | $R$ | Universal gas constant, (J/(mol.K)) |
| $A_s$ | Average area of the dissociation front | $R_1$ | Thermal resistivity for heat transfer through inner tube, (m.K/W) |
| $A_w$ | Wellbore area, (m$^2$) | $R_2$ | Thermal resistivity for heat transfer through annulus, (m.K/W) |
| $a$ | Dimensionless constant in equation 37 | $R_w$ | Thermal resistivity of the wellbore, (m.K/W) |
| $B$ | Dimensionless variable in equation 37 | $R_{cia}$ | Convection thermal resistance between water flow and outer surface of the annulus, (m.K/W) |
| $B_a$ | Dimensionless constant in equation 24 | $R_{cii}$ | Convective heat transfer resistance between the inner tube surface and the fluid flow in the tube, (m.K/W) |
| BC 1 | Boundary conditions | $R_{coi}$ | Convection thermal resistance between the inner tube and the fluid flow in the annulus, (m.K/W) |
| BC 2 | Boundary conditions | $R_p$ | Conduction thermal resistance of the inner tube, (m.K/W) |
| $b$ | Dimensionless constant in equation 37 | $r$ | Radial distance, (m) |
| $C$ | Dimensionless variable in equation 38 | $r_1$ | Inside radius of inner tube, (m) |
| $C_f$ | Fluid volumetric heat capacity, (kJ/(kg.K)) | $r_2$ | Outside radius of inner tube, (m) |
| $C_{pI}$ | Specific heat capacity of Zone I, (J/(kg.K)) | $r_3$ | Inside radius of wellbore/ Inside radius of casing 1, (m) |
| $C_{pg}$ | Specific heat capacity of gas, (J/(kg.K)) | $r_4$ | Outside radius of casing 1/ Inside radius of gravel part, (m) |
| $c$ | Dimensionless constant in equation 27 | $r_5$ | Outside radius of gravel part/ Inside radius of casing 2, (m) |
| $D$ | Dimensionless variable in equation S14 | $r_6$ | Outside radius of casing 2/ Inside radius of cement part, (m) |
| $d$ | Dimensionless constant in equation 27 | $r_7$ | Outside radius of cement part/ Outside radius of wellbore, (m) |
| $d_h$ | Hydraulic diameter of the annulus, (m) | $r^*$ | Ratio of the inner and outer radii of the annulus $(r_2 / r_3)$ |
| $E$ | Dimensionless variable in equation S14 | $\mathrm{Re}_i$ | Reynolds number in the inner tube |
| $F$ | Dimensionless variable in equation S14 | $S$ | Interface position, (m) |
| $F_{gH}$ | Ratio of mass of the methane gas trapped inside the MH to the mass of hydrate | $S_t$ | Interface position at time $t$, (m) |
| $f$ | Friction factor | $T_{fi}$ | Water flow temperature in the inner tube, (K) |
| $G(\beta)$ | Dimensionless constant in equation S24 | $T_{fa}$ | Water flow temperature in the annulus, (K) |
| $H(\beta)$ | Dimensionless constant in equation S24 | $T_i$ | Hot water injection temperature, (K) |
| $h$ | Thickness of MH reservoir, (m) | $T_I$ | Temperature in Zone I, (K) |
| $I(\beta)$ | Dimensionless constant in equation S24 | $T_{II}$ | Temperature in Zone II, (K) |
| $K(\beta)$ | Function in equation 40 | $T_0$ | Initial temperature of hydrate, (K) |
| $k$ | Permeability, (μm$^2$) | $T_{fi}$ | Fluid temperature in the inner tube, (K) |
| $k_f$ | Fluid thermal conductivity, (W/(m.K)) | $T_{fa}$ | Fluid temperature in the annulus, (K) |
| $k_p$ | Inner tube's thermal conductivity, (W/(m.K)) | $T_s$ | Temperature at the interface, (K) |
| $k_I$ | Thermal conductivity of Zone I, (W/(m.K)) | $T_{STP}$ | Temperature of gas at STP conditions, (K) |
| $k_{II}$ | Thermal conductivity of Zone II, (W/(m.K)) | $t$ | Time, (s) |
| $k_c$ | Thermal conductivity of cement, (W/(m.K)) | $u_r$ | Heat flux from the well, (J/(m$^2$. s)) |
| $k_g$ | Thermal conductivity of gravel, (W/(m.K)) | $V_f$ | Water flow rate, (m$^3$/s) |
| $k_p$ | Thermal conductivity of inner tube, (W/(m.K)) | $V_g$ | Volume of produced gas per surface area of the moving interface in the time fraction of "*t,t-1*", (m$^3$/m$^2$) |
| $k_s$ | Thermal conductivity of casing, (W/(m.K)) | $V_{rp}$ | Total volume of produced gas per surface area of the moving interface up to time *t*, (m$^3$/m$^2$) |
| $L(\beta)$ | Function in equation 40 | $v_s$ | Interface velocity, (m/s) |
| $M(\beta)$ | Function in equation S19 | $v_f$ | Water flow velocity, (m/s) |
| MH | Methane hydrate | $v_g$ | Gas velocity, (m/s) |
| $m$ | Gas molecular mass, (g/mol) | $z$ | Axial distance along the wellbore, (m) |
| $N(\lambda)$ | Function in equation 40 | $\rho_I$ | Density of the matrix in Zone I, (kg/m$^3$) |
| $Nu_{ii}$ | Nusselt's number of the fluid flow in the inner tube | $\rho_f$ | Fluid density, (kg/m$^3$) |
| $Nu_{oa}$ | Nusselt's number of the water flow close to the outer surface of annulus | $\rho_g$ | Gas density, (kg/m$^3$) |
| $Nu_{oi}$ | Nusselt's number of the fluid flow in the annulus close to the inner tube | $\rho_H$ | Hydrate density, (kg/m$^3$) |
| $Nu_{pipe}$ | Nusselt's number of the water flow in the annulus close to the inner pipe | $\mu$ | Gas viscosity, (kg/m$^3$) |
| $n_r$ | Total moles of produced gas per surface area of the moving interface in the time fraction of "*t,t-1*", (mole/m$^2$) | $\lambda$ | Dimensionless variable in equation 34 |
| $n_{rt}$ | Total moles of produced gas per surface area of the moving interface up to time *t*, (mole/m$^2$) | $\lambda_{os}$ | Dimensionless variable in equation 36 |
| $P$ | Pressure in Zone I, (Pa) | $\beta$ | Dimensionless constant in equation 35 |
| $P_s$ | Pressure at the interface, (Pa) | $\phi$ | Porosity |
| $P_i$ | Pressure of the heat source, (Pa) | $\alpha_I$ | Thermal diffusivity of Zone I, (μm$^2$/s) |
| $P_{STP}$ | Pressure of gas at STP conditions, (Pa) | $\alpha_{II}$ | Thermal diffusivity of Zone II, (μm$^2$/s) |
| $\Pr$ | Prandtl number | $\mu_f$ | Fluid dynamic viscosity, (Pa.s) |
| $Q_{Hd}$ | Heat of MH dissociation, (J/kg) | $\nu_f$ | Water kinematic viscosity, (m$^2$/s) |
| $Q_g$ | Heating value of the gas at STP conditions (MJ/m$^3$) | $\eta_r$ | energy efficiency ratio |
| $Q_{rt}$ | Total input heat to the reservoir from the heat source, (J/m$^2$) | $\zeta$ | Constant in equation 8 |
| $q_a$ | Heat transfer rate from annulus to the reservoir, (MJ/(m$^2$.s)) | | |

## Supplementary Information

- **Materials and methods**

In the following, the solution process of equations 16 and 17 is provided. This process is a little bit different for the two operation models. For the model in which the hot water is injected to the inner tube, $T_{fa}$ is calculated first based on $T_{fi}$ as follows:

$$T_{fa}(z,t) = R_1 C_f V_f \frac{\partial T_{fi}(z,t)}{\partial z} + T_{fi}(z,t) \tag{S1}$$

Then, by inserting equation S1 to equation 17, the resulted expression is:

$$R_1 R_2 (C_f V_f)^2 \frac{\partial^2 T_{fi}(z,t)}{\partial z^2} - R_1 C_f V_f \frac{\partial T_{fi}(z,t)}{\partial z} - T_{fi}(z,t) + T_I(r_7,t) \tag{S2}$$

Equation S2 is a second-order nonhomogeneous differential equation, which can be solved to obtain $T_{fi}$ due to the following initial and boundary conditions: i) injection temperature is fixed and constant through the process ($T_{fi}(h,t)$); and ii) no heat flow occurs at the base of the wellbore ($T_{fi}(0,t) = T_{fa}(0,t) \rightarrow \frac{\partial T_{fi}(0,t)}{\partial z} = \frac{\partial T_{fa}(0,t)}{\partial z} = 0$).

For the other model of how water injection into the annulus, $T_{fa}$ based on $T_{fi}$ is as follows:

$$T_{fa}(z,t) = R_1 C_f V_f \frac{\partial T_{fi}(z,t)}{\partial z} + \frac{R_1}{R_2}(T_{fi}(z,t) - T_I(r_7,t)) + T_{fi}(z,t) \tag{S3}$$

Then, by inserting equation S3 to equation 17, the new expression based on $T_{fi}$ is:

$$R_1 R_2 (C_f V_f)^2 \frac{\partial^2 T_{fi}(z,t)}{\partial z^2} + R_1 C_f V_f \frac{\partial T_{fi}(z,t)}{\partial z} - T_{fi}(z,t) + T_I(r_7,t) \tag{S4}$$

Equation S4 is also a second-order nonhomogeneous differential equation, and can be solved by the same procedure and with the same initial and boundary conditions as those of equation S2.

In the following, the transformation of the fundamental equations of dissociation process with the initial and boundary conditions in terms of $\lambda$ is provided.

Equation 21: $\frac{d^2 T_{II}}{d\lambda^2} + \left(\frac{2\lambda^2+1}{\lambda}\right)\frac{dT_{II}}{d\lambda} = 0$, $\beta < \lambda < \infty$ (S5)

Equation 23: $\frac{-k_I A_w}{\sqrt{4\alpha_{II} t}} \frac{dT_I}{d\lambda} = \frac{(\overline{T_{fa}(t)} - T_I)}{R_w}$, $\lambda = \lambda_{os}$ (S6)

Equation 24: $P_s(t) = \exp(A_a - B_a / T_s(t))$, $\lambda = \beta$ (S7)

Equation 26: $k_{II}\frac{dT_{II}}{d\lambda} - k_I \frac{dT_I}{d\lambda} = 2\phi\rho_H \alpha_{II} Q_{Hd}\beta$, $\lambda = \beta$ (S8)

Equation 31: $\frac{d}{d\lambda}\left(\frac{P}{T_I}\frac{dP}{d\lambda}\right) + \frac{2\phi\alpha_{II}\mu\lambda}{k}\frac{d}{d\lambda}\left(\frac{P}{T_I}\right) = 0$, $\lambda_{os} < \lambda < \beta$ (S9)

Equation 32: $\frac{d^2 T_I}{d\lambda^2} + \left(\frac{2\alpha_{II}\lambda}{\alpha_I} + \frac{1}{\lambda}\right)\frac{dT_I}{d\lambda} + \frac{C_{pg} km}{2k_I \mu R}\frac{d^2 P^2}{d\lambda^2} = 0$, $\lambda_{os} < \lambda < \beta$ (S10)

Equation 33: $\frac{P}{T_I}\frac{dP}{d\lambda} = \frac{2F_{gH}\phi\rho_H \alpha_{II}\mu R\beta}{km}$, $\lambda = \beta$ (S11)

And, from equation 29: $T_{II} = T_0$, $\lambda \to \infty$ (S12)

Selim et al. (Selim and Sloan 1990) reported that the very small ratio of $\frac{\rho_g}{\rho_H}$ indicates the very slow rate of hydrate dissociation (dS/dt), which neglects the transient term in the continuity equation (second term in equation S5).

In the equations 37 and 38, $A$, $A_1$, $B$, $B_1$, and $C$ constants are defined using equations S2 and 14 as follows:

$$A = \frac{(T_s - \overline{T_{fa}(t)})F}{-FEi(-(a\beta + b)^2) + FEi(-b^2) + D + E} \tag{S13}$$

$$B = \frac{A(D+E)}{F} + \overline{T_{fa}(t)} \tag{S14}$$

$$C = \frac{T_s(t) - T_0}{Ei(-\beta^2)} \tag{S15}$$

, where *a*, *b*, *D*, *E*, and *F* are as follows:

$$a = \left(\frac{\alpha_{II}}{\alpha_I}\right)^{1/2} \tag{S16}$$

$$b = \frac{C_{pg} F_{gH} \phi \rho_H \alpha_{II} \beta}{a k_I} \tag{S17}$$

$$D = \frac{-r_7 k_I 2a \exp(-(a\lambda_{os} + b)^2)}{e\sqrt{\pi}} \tag{S18}$$

$$E = R_w(-Ei(-(a\lambda_{os} + b)^2) + Ei(-b^2)) \tag{S19}$$

$$F = \frac{1}{R_w} \tag{S20}$$

The $L(\beta)$, $M(\beta)$, $N(\lambda)$, and $K(\beta)$ functions used in equation 40 (pressure distribution in zone I) are as follows:

$$L(\beta) = \frac{4F_{gH} \phi \rho_H \alpha_{II} \mu R \beta}{km} \tag{S21}$$

$$M(\beta) = \overline{T_{fa}(t)} + AEi(-b^2) \tag{S22}$$

$$N(\lambda) = \frac{Ei(-(a\lambda + b)^2)(a\lambda + b) - \sqrt{\pi} erf(a\lambda + b)}{a} \tag{S23}$$

$$K(\beta) = B + AEi(-b^2) \tag{S24}$$

Now, equation S8 by insertion of the resulted $T_I$ and $T_{II}$ formulas and considering equation 13 becomes:

$$\frac{ak_I(\overline{T_{fa}(t)}-T_s(t))}{k_{II}(T_s(t)-T_0)}\underbrace{\left(\frac{-F\exp(-(a\beta+b)^2)}{(a\beta+b)(-FEi(-(a\beta+b)^2)+FEi(-b^2)-D-E)}\right)}_{G(\beta)}+\underbrace{\left(\frac{\exp(-\beta^2)}{\beta Ei(-\beta^2)}\right)}_{H(\beta)}=\underbrace{\sqrt{\pi}\phi\rho_H\alpha_{II}\beta}_{I(\beta)}\frac{(c+dT_s(t))}{k_{II}(T_s(t)-T_0)} \quad \text{(S25)}$$

So, the temperature at the dissociation interface based on equation S23 can be calculated as follows:

$$T_s(t)=\frac{k_{II}T_0H(\beta)-ak_I\overline{T_{fa}(t)}G(\beta)+cI(\beta)}{-ak_IG(\beta)-dI(\beta)+k_{II}H(\beta)} \quad \text{(S26)}$$

The pressure at the dissociation interface ($P_s$) would be calculated from the equation 40:

$$P_s(t)=\left(P_i^2+L(\beta)(K(\beta)\beta-AN(\beta)-(K(\beta)\lambda_{os}-AN(\lambda_{os})))\right)^{1/2} \quad \text{(S27)}$$

The following equation is a transformation of equation 24 (Antoine equation):

$$\frac{B_a}{A_a-\ln(P_s)}-T_s=0 \quad \text{(S28)}$$

Then, by replacing $T_s$ and $P_s$ in equation S28 with the equations S26 and S27, and solving the resulted expression to find $\beta$, the exact solution of temperature and pressure distributions will be achieved. Actually, at each time step, equation S28 has only one unknown, which is $\beta$, and MATLAB programming software is used to find this unknown as well as performing all other calculations in this study.

The transformed form of equation of heat flux from the wellbore (equation 41) based on the transformation term of equation 34 is:

$$u_r=\frac{-k_I}{\sqrt{4\alpha_{II}t}}\frac{dT_I}{d\lambda},\ \lambda=\lambda_{os} \quad \text{(S29)}$$

By inserting $T_I$ from equation 37 into equation S29:

$$u_r=\frac{2ak_IA\exp(-(a\lambda_{os}+b)^2)}{(a\lambda_{os}+b)\sqrt{4\alpha_{II}t}} \quad \text{(S30)}$$

In order to obtain the total volume of gas produced at STP conditions, the produced gas moles at each time step should be calculated. Equations S32 and S33 show respectively the volume of dissociated hydrate and the moles of produced gas per average surface area of the dissociation front at each time step (*t*-1, *t*) as shown in equation S31.

$$A_s = 2\pi(\frac{S_t + S_{t-1}}{2}) \tag{S31}$$

$$V_{rp} = \phi h \pi (S_t^2 - S_{t-1}^2) \tag{S32}$$

$$n_r = F_{gH} \rho_H V_{rp} / (A_s m) \tag{S33}$$

, where *h* is the methane hydrate thickness (m). By summing the moles of produced gas at each time step from the beginning of dissociation to time *t* as shown in equation S34, the total number of moles of produced gas up to that time can be calculated.

$$n_{rt} = \sum_{t=0}^{t} n_r \tag{S34}$$

Finally, by using equations S34 and 36 the total volume of produced gas per average area of the dissociation front at STP conditions up to time *t* can be achieved as presented in equation 43.

---

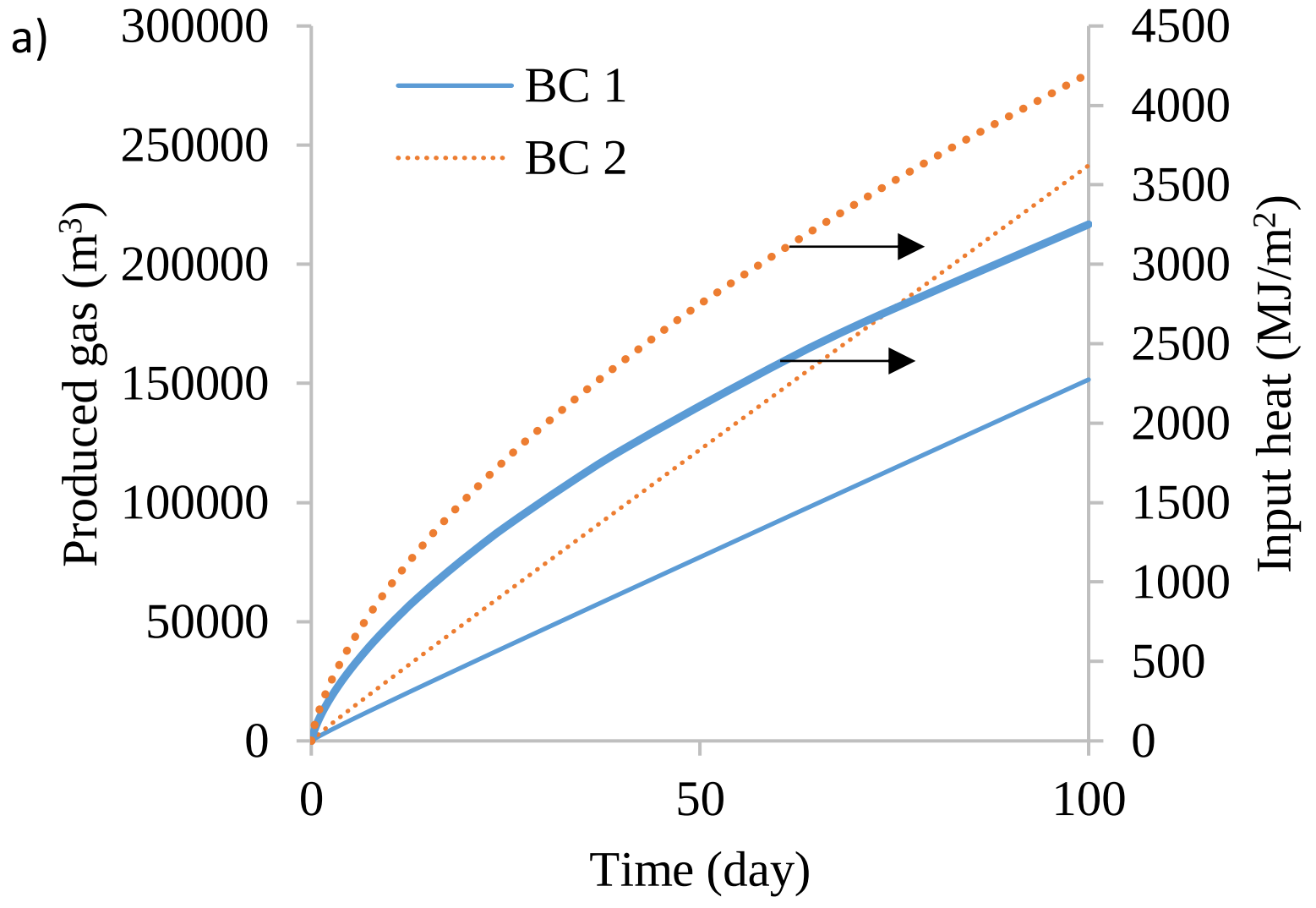

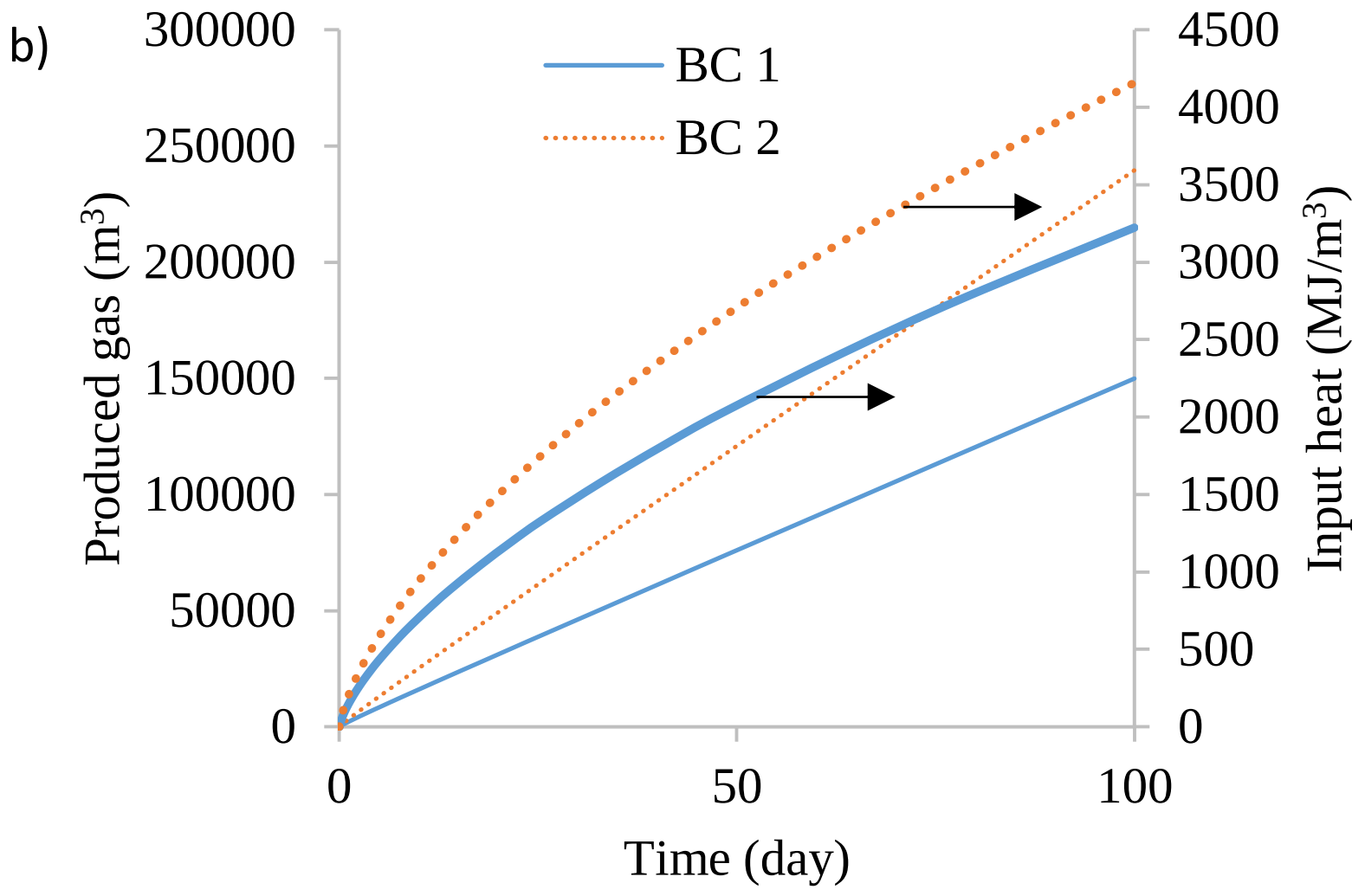

Figure S1. Volume of produced gas and the input heat from the wellbore in the model with hot water injection into the a) annulus and b) inner tube during hydrate dissociation for two BCs. The thicker lines represent the input heat from the wellbore.

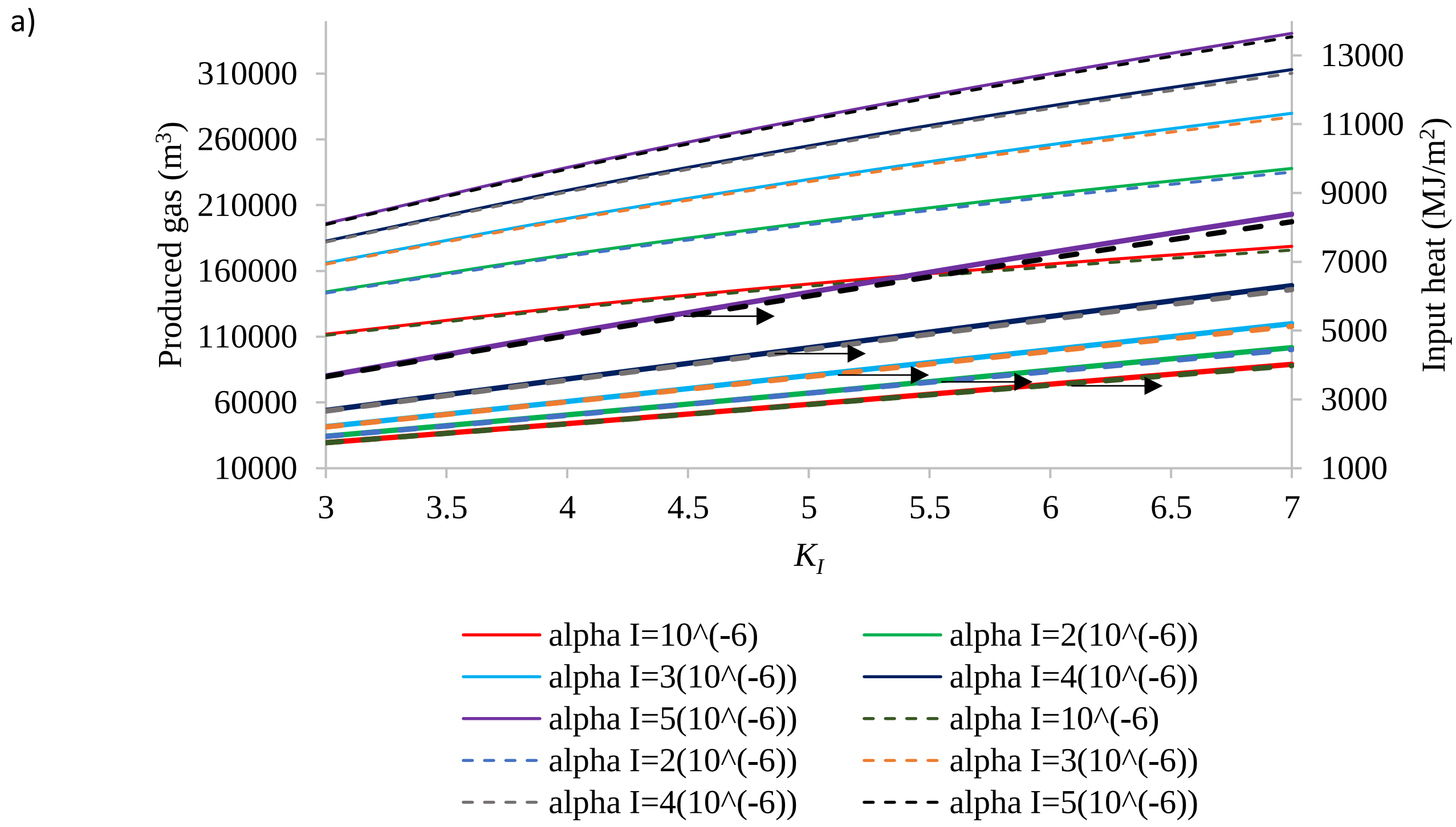

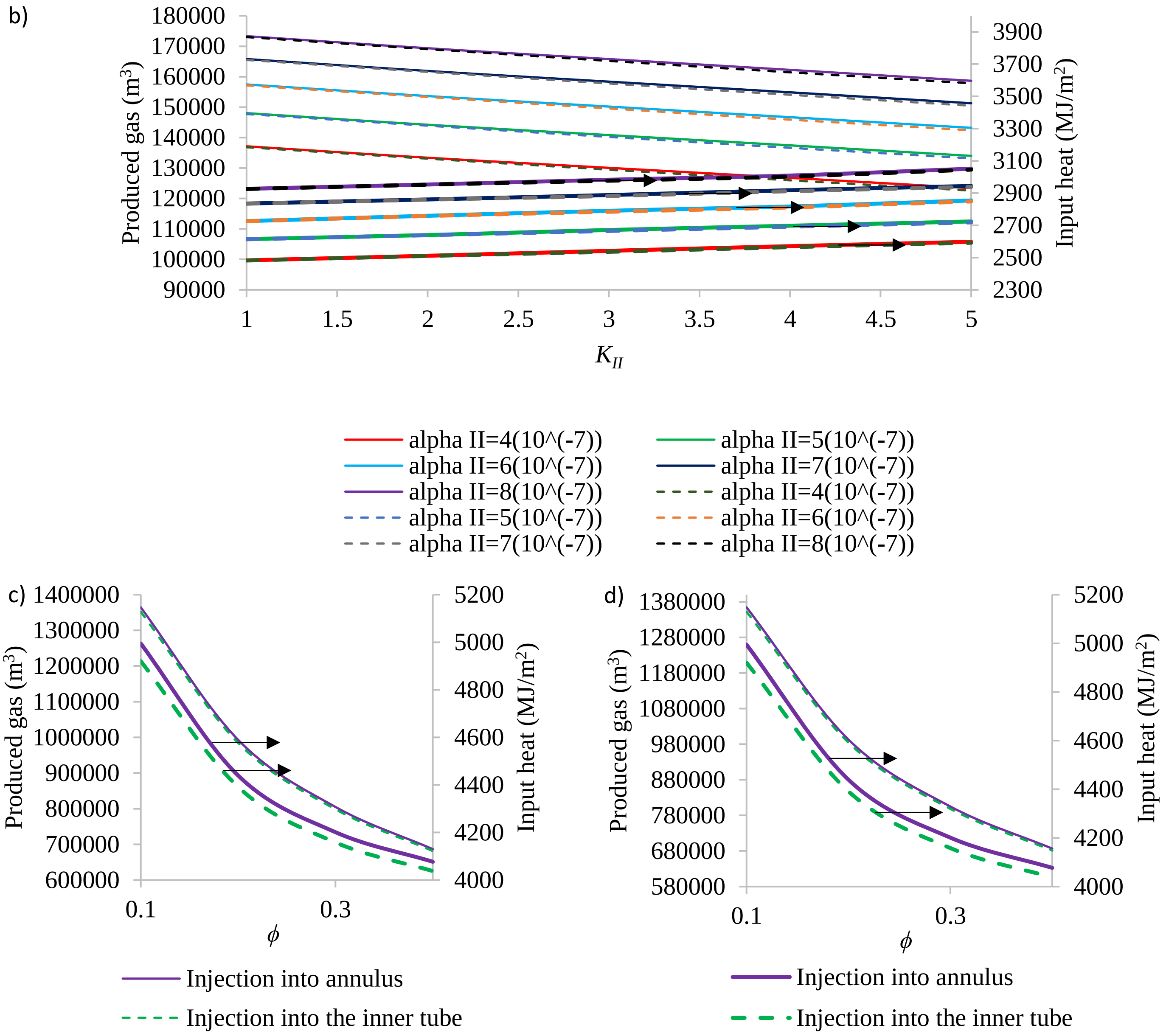

Figure S2. The effect of various parameters on the produced gas and the input heat from the wellbore after 100 days dissociation considering both operating schemes: a) thermal diffusivity and thermal conductivity of Zone I, b) thermal diffusivity and thermal conductivity of Zone II, c) porosity with various permeabilities, and d) porosity with various gas viscosities.

Dashed lines and solid lines are respectively representative of the operating schemes of hot water injection into annulus and into the inner tube. The thicker lines represent the input heat from the wellbore.

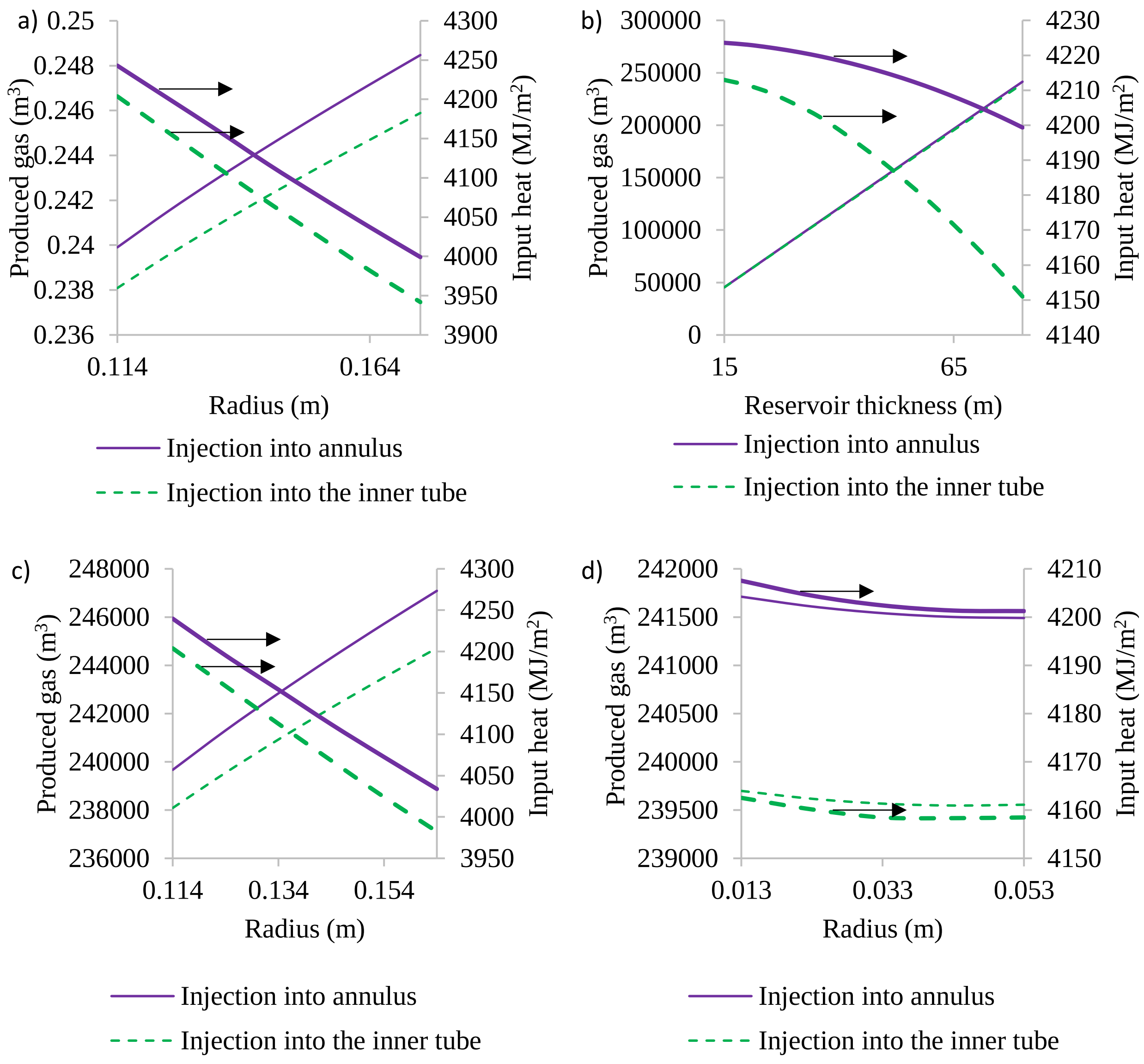

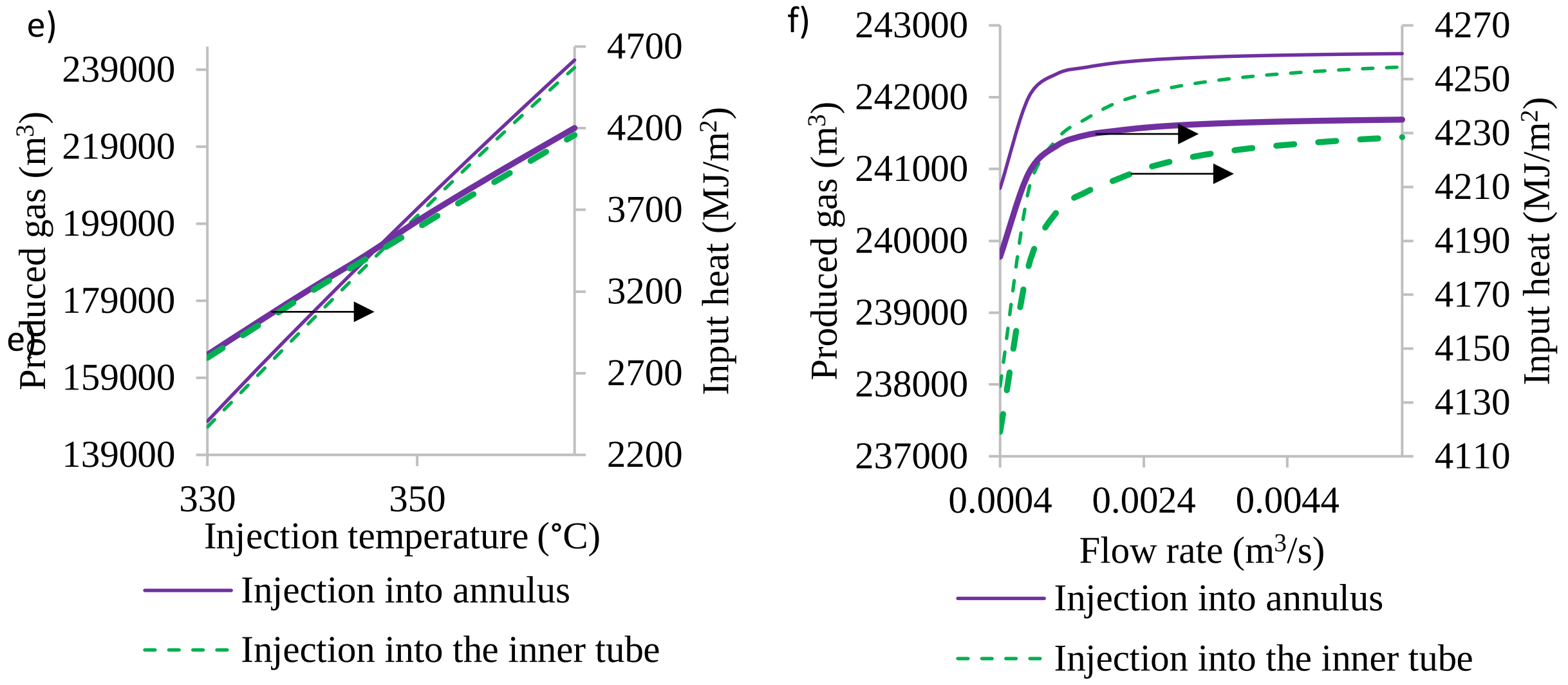

Figure S3. The effect of various parameters on the produced gas and the input heat from the wellbore after 100 days dissociation considering both operating schemes: a) the wellbore radius, b) the reservoir thickness, c) the annulus radius, d) the inner tube radius, e) the injection temperature, and f) the flow rate.

Dashed lines and solid lines are respectively representative of the operating schemes of hot water injection into annulus and into the inner tube. The thicker lines represent the input heat from the wellbore.